\documentclass[acmsmall,authorversion]{acmart}

\acmConference[ICFP'17]{ACM SIGPLAN International Conference on Functional Programming}{September 04--06, 2017}{Oxford, UK}
\acmJournal{PACMPL}
\acmYear{2017}
\acmISBN{978-x-xxxx-xxxx-x/YY/MM}
\acmDOI{10.1145/nnnnnnn.nnnnnnn}
\startPage{1}

\usepackage[utf8]{inputenc}
\usepackage[color=green!40]{todonotes}
\usepackage{paralist}
\usepackage{hyperref}
\usepackage[capitalise,nameinlink,noabbrev]{cleveref}
\usepackage{upgreek}
\usepackage{tikz}
\usetikzlibrary{calc}
\usepackage{keystroke}
\usepackage{framed}
\usepackage{adjustbox}

\newcommand{\lmss}{\fontencoding{T1}\fontfamily{lmss}\selectfont} 

\usepackage{wrapfig}
\makeatletter
\@ifundefined{ifediting}{
\newif\ifediting
\editingfalse
}{}
\makeatother

\makeatletter
\@ifundefined{iflong}{
\newif\iflong
\longtrue
}{}
\makeatother

\ifediting
\paperwidth=\dimexpr \paperwidth + 6cm\relax
\oddsidemargin=\dimexpr\oddsidemargin + 0cm\relax
\evensidemargin=\dimexpr\evensidemargin + 6cm\relax
\marginparwidth=\dimexpr \marginparwidth + 3cm\relax

\newcommand{\jb}[2][]{\todo[color=green!40, #1]{#2}}
\newcommand{\cs}[2][]{\todo[color=blue!20, #1]{#2}}
\else
\newcommand{\jb}[2][]{}
\newcommand{\cs}[2][]{}
\fi

\definecolor{P1}{named}{green}
\definecolor{P2}{named}{blue}

\makeatletter
\@ifundefined{lhs2tex.lhs2tex.sty.read}%
  {\@namedef{lhs2tex.lhs2tex.sty.read}{}%
   \newcommand\SkipToFmtEnd{}%
   \newcommand\EndFmtInput{}%
   \long\def\SkipToFmtEnd#1\EndFmtInput{}%
  }\SkipToFmtEnd

\newcommand\ReadOnlyOnce[1]{\@ifundefined{#1}{\@namedef{#1}{}}\SkipToFmtEnd}
\usepackage{amstext}
\usepackage{amssymb}
\usepackage{stmaryrd}
\DeclareFontFamily{OT1}{cmtex}{}
\DeclareFontShape{OT1}{cmtex}{m}{n}
  {<5><6><7><8>cmtex8
   <9>cmtex9
   <10><10.95><12><14.4><17.28><20.74><24.88>cmtex10}{}
\DeclareFontShape{OT1}{cmtex}{m}{it}
  {<-> ssub * cmtt/m/it}{}

\DeclareFontShape{OT1}{cmtt}{bx}{n}
  {<5><6><7><8>cmtt8
   <9>cmbtt9
   <10><10.95><12><14.4><17.28><20.74><24.88>cmbtt10}{}
\DeclareFontShape{OT1}{cmtex}{bx}{n}
  {<-> ssub * cmtt/bx/n}{}

\newcommand{\Conid}[1]{\mathit{#1}}
\newcommand{\Varid}[1]{\mathit{#1}}
\newcommand{\anonymous}{\kern0.06em \vbox{\hrule\@width.5em}}

\newcommand{\bind}{\mathbin{>\!\!\!>\mkern-6.7mu=}}
\newcommand{\sequ}{\mathbin{>\!\!\!>}}
\renewcommand{\leq}{\leqslant}
\renewcommand{\geq}{\geqslant}
\usepackage{polytable}

\@ifundefined{mathindent}%
  {\newdimen\mathindent\mathindent\leftmargini}%
  {}%

\def\resethooks{%
  \global\let\SaveRestoreHook\empty
  \global\let\ColumnHook\empty}
\newcommand*{\savecolumns}[1][default]%
  {\g@addto@macro\SaveRestoreHook{\savecolumns[#1]}}
\newcommand*{\restorecolumns}[1][default]%
  {\g@addto@macro\SaveRestoreHook{\restorecolumns[#1]}}
\newcommand*{\aligncolumn}[2]%
  {\g@addto@macro\ColumnHook{\column{#1}{#2}}}

\resethooks

\newcommand{\onelinecommentchars}{\quad-{}- }
\newcommand{\commentbeginchars}{\enskip\{-}
\newcommand{\commentendchars}{-\}\enskip}

\newcommand{\visiblecomments}{%
  \let\onelinecomment=\onelinecommentchars
  \let\commentbegin=\commentbeginchars
  \let\commentend=\commentendchars}

\newcommand{\invisiblecomments}{%
  \let\onelinecomment=\empty
  \let\commentbegin=\empty
  \let\commentend=\empty}

\visiblecomments

\newlength{\blanklineskip}
\newcommand{\hsindent}[1]{\quad}
\let\hspre\empty
\let\hspost\empty

\EndFmtInput
\makeatother
\ReadOnlyOnce{polycode.fmt}%
\makeatletter

\newcommand{\hsnewpar}[1]%
  {{\parskip=0pt\parindent=0pt\par\vskip #1\noindent}}

\newcommand{\hscodestyle}{}

\newcommand{\sethscode}[1]%
  {\expandafter\let\expandafter\hscode\csname #1\endcsname
   \expandafter\let\expandafter\endhscode\csname end#1\endcsname}

  {\par\noindent
   \advance\leftskip\mathindent
   \hscodestyle
   \let\\=\@normalcr
   \let\hspre\(\let\hspost\)%
   \pboxed}%
  {\endpboxed\)%
   \par\noindent
   \ignorespacesafterend}

  {\hsnewpar\abovedisplayskip
   \advance\leftskip\mathindent
   \hscodestyle
   \let\hspre\(\let\hspost\)%
   \pboxed}%
  {\endpboxed%
   \hsnewpar\belowdisplayskip
   \ignorespacesafterend}

  {\hsnewpar\abovedisplayskip
   \advance\leftskip\mathindent
   \hscodestyle
   \let\\=\@normalcr
   \(\pboxed}%
  {\endpboxed\)%
   \hsnewpar\belowdisplayskip
   \ignorespacesafterend}

\newcommand{\plainhs}{\sethscode{plainhscode}}

\plainhs

  {\hsnewpar\abovedisplayskip
   \advance\leftskip\mathindent
   \hscodestyle
   \let\\=\@normalcr
   \(\parray}%
  {\endparray\)%
   \hsnewpar\belowdisplayskip
   \ignorespacesafterend}

  {\parray}{\endparray}

  {\(\parray}{\endparray\)}

\def\codeframewidth{\arrayrulewidth}
\RequirePackage{calc}

  {\parskip=\abovedisplayskip\par\noindent
   \hscodestyle
   \arrayrulewidth=\codeframewidth
   \tabular{@{}|p{\linewidth-2\arraycolsep-2\arrayrulewidth-2pt}|@{}}%
   \hline\framedhslinecorrect\\{-1.5ex}%
   \let\endoflinesave=\\
   \let\\=\@normalcr
   \(\pboxed}%
  {\endpboxed\)%
   \framedhslinecorrect\endoflinesave{.5ex}\hline
   \endtabular
   \parskip=\belowdisplayskip\par\noindent
   \ignorespacesafterend}

\newcommand{\framedhslinecorrect}[2]%
  {#1[#2]}

  {\(\def\column##1##2{}%
   \let\>\undefined\let\<\undefined\let\\\undefined
   \newcommand\>[1][]{}\newcommand\<[1][]{}\newcommand\\[1][]{}%
   \def\fromto##1##2##3{##3}%
   }{\) }%

  {\let\orighscode=\hscode
   \let\origendhscode=\endhscode
   \def\endhscode{\def\hscode{\endgroup\def\@currenvir{hscode}\\}\begingroup}
   \orighscode\def\hscode{\endgroup\def\@currenvir{hscode}}}%
  {\origendhscode
   \global\let\hscode=\orighscode
   \global\let\endhscode=\origendhscode}%

\makeatother
\EndFmtInput
  {\hsnewpar\abovedisplayshortskip
   \advance\leftskip\mathindent
   \hscodestyle
   \let\hspre\(\let\hspost\)%
   \pboxed}%
  {\endpboxed%
   \hsnewpar\belowdisplayshortskip
   \ignorespacesafterend}

\sethscode{myhscode}

\ReadOnlyOnce{forall.fmt}%
\makeatletter

\let\HaskellResetHook\empty
\newcommand*{\AtHaskellReset}[1]{%
  \g@addto@macro\HaskellResetHook{#1}}
\newcommand*{\HaskellReset}{\HaskellResetHook}

\newcommand\hsforall{\global\let\hsdot=\hsperiodonce}
\newcommand*\hsperiodonce[2]{#2\global\let\hsdot=\hscompose}
\newcommand*\hscompose[2]{#1}

\AtHaskellReset{\global\let\hsdot=\hscompose}

\HaskellReset

\makeatother
\EndFmtInput
\ReadOnlyOnce{lambda.fmt}%
\makeatletter

\newcommand\hslambda{\global\let\hsarrow=\hsarrowperiodonce}
\newcommand*\hsarrowperiodonce[2]{#2\global\let\hsarrow=\hscompose}

\AtHaskellReset{\global\let\hsarrow=\hscompose}

\HaskellReset

\makeatother
\EndFmtInput
\renewcommand{\Varid}[1]{\textsf{\lmss #1}}
\renewcommand{\Conid}[1]{\textsf{\lmss #1}}

\renewcommand{\onelinecomment}{\,--\hspace{.5pt}--\itshape{} }
\begin{document}
\title{Lock-step simulation is child's play}
\subtitle{\iflong Experience Report -- Extended Version\else Experience Report\fi}

\author{Joachim Breitner}
\affiliation{%
  \institution{University of Pennsylvania}
  \streetaddress{3330 Walnut Street}
  \city{Philadelphia}
  \state{PA}
  \postcode{19104}
  \country{USA}}
\email{joachim@cis.upenn.edu}

\author{Chris Smith}
\affiliation{Google}

\begin{abstract}
Implementing multi-player networked games by broadcasting the player's input and letting each client calculate the game state -- a scheme known as \emph{lock-step simulation} -- is an established technique. However, ensuring that every client in this scheme obtains a consistent state is infamously hard and in general requires great discipline from the game programmer. The thesis of this report is that in the realm of functional programming -- in particular with Haskell's purity and static pointers -- this hard problem becomes almost trivially easy.

We support this thesis by implementing lock-step simulation under very adverse conditions. We extended the educational programming environment CodeWorld, which is used to teach math and programming to middle school students, with the ability to create and run interactive, networked multi-user games. Despite providing a very abstract and high-level interface, and without requiring any discipline from the programmer, we can provide consistent lock-step simulation with client prediction.
\end{abstract}

\maketitle

Networked multi-user games must tackle the challenge of ensuring that all participating players on a network with potentially significant latency still see the same game state. In some circumstances, an appealing choice is \emph{lock-step simulation}. In this scheme, which dates back to the age of Doom, the state of the game itself is never transmitted over the network. Instead, the clients exchange information about their player's interactions -- as abstract game moves or just the actual user input events -- and each client independently calculates the state of the game.

Of course, this only works as intended if all clients end up with the same state. The technique is fraught with danger if the programmer is not very careful and disciplined about managing that state. \citet{terranobettner}, who implemented the network code for the real time strategy games Age of Empires 1 \& 2,  report:
\begin{quote}
As much as we check-summed the world, the objects, the pathfinding, targeting and every other system -- it seemed that there was always one more thing that slipped just under the radar. [\ldots] Part of the difficulty was conceptual -- programmers were not used to having to write code that used the same number of calls to random within the simulation.
\end{quote}

\goodbreak
More drastic words were voiced by \citet{forrestsmith}, also a video game software engineer:
\begin{quote}
One of the most vile bugs in the universe is the desync bug. They're mean sons of bitches. The grand assumption to the entire engine architecture is all players being fully synchronous. What happens if they aren't? What if the simulations diverge? Chaos. Anger. Suffering.
\end{quote}

\begin{wrapfigure}[16]{r}{0.4\linewidth}%
\fbox{\includegraphics[width=\linewidth]{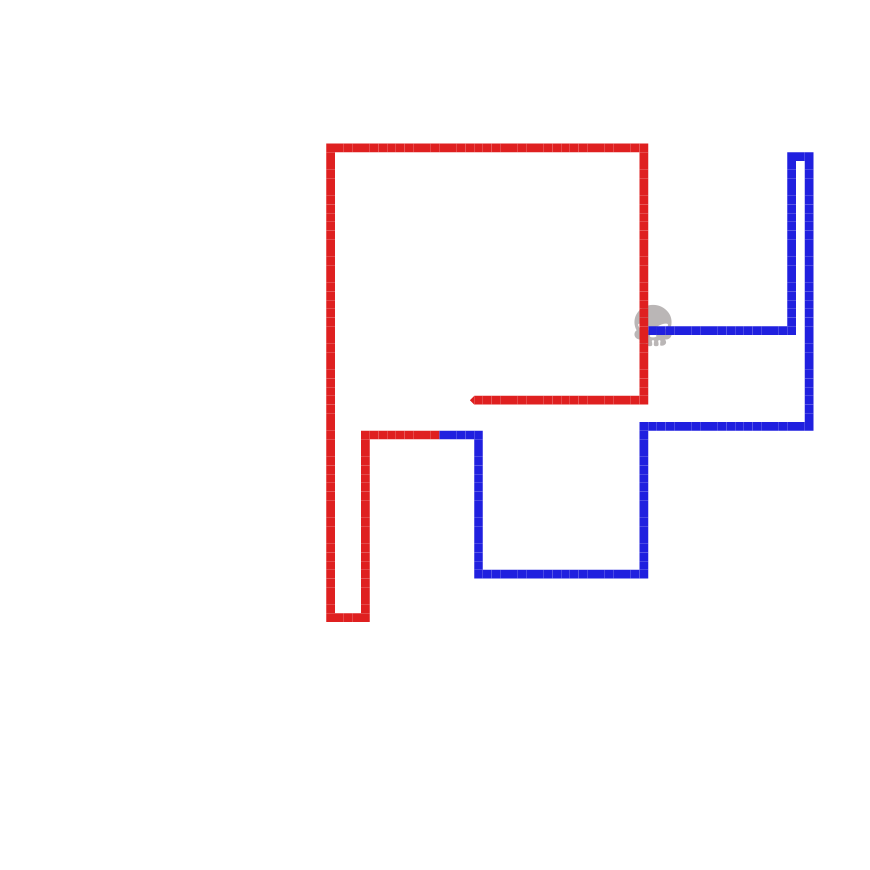}}
\caption{The Snake game}
\label{fig:snake}
\end{wrapfigure}

The pitfalls facing a programmer implementing lock-step simulation include reading the system clock, querying the random number generator, other I/O, uninitialized memory, and local or hidden statefulness. In short: side effects! What if we chose a programming language without such side-effects? Would these problems disappear? Intuitively, we expect that \textbf{pure functional programming makes lock-step simulation easy}.

This experience report corroborates our expectation. We have implemented lock-step simulation in Haskell under very adverse conditions. The authors of the quotes above are professional programmers working on notable games. They can be expected to maintain a certain level of programming discipline, and to tolerate additional complexity. Our implementation is part of CodeWorld\footnote{\url{https://code.world/haskell}}, an educational, web-based programming environment used to teach mathematics and coding to students as early as middle school. These children, who are just learning to code, can write and run arbitrary game logic, using a simple API, without adhering to any additional requirements or coding discipline. Nevertheless, we still guarantee consistent lock-step simulation and avoid the dreaded desync bug.

The main contributions of this experience report are:
\begin{itemize}
\item With a bold disregard for pesky implementation detail, we design a natural extension to CodeWorld's existing interfaces that can describe multi-user interactive programs in as straightforward, simple and functional a manner as possible (\cref{sec:wishful}).
\item We identify a complication -- unwanted capture of free variables -- which can thwart consistency of such a program.  We solve it using either using the module system (\cref{sec:no-io}) or the Haskell language extension \emph{static pointers} (\cref{sec:static-pointers}).
\item We explain how to implement this interface. Despite its abstractness, we present an eventually consistent implementation that works for arbitrary client code, and includes \emph{client prediction} to react immediately to local input while still reconciling delayed input from other users (\cref{sec:prediction}).
\iflong
\item We share lessons learned in stress-testing the system (\cref{sec:experience}). Testing was successful, but we identified an inconsistency in floating point transcendental functions. Replacing these with deterministic approximations recovers the consistency that we rely upon (\cref{sec:floatingpoint}).
\else
\item Haskell’s promise of purity is muddied by underspecification of floating point transcendental functions. Replacing these with deterministic approximations recovers the consistency that we rely upon (\cref{sec:floatingpoint}).
\fi
\item We show that, even with no knowledge of the structure of the program's state, our approach still allows us to smooth out artifacts that arise due to network latency (\cref{sec:smoothing}).
\item Overall, we show that pure functional programming makes lock-step simulation easy.
\end{itemize}

\iflong
\pagebreak
\fi

\section{CodeWorld}

In this section, we give a brief overview of how students interact with the CodeWorld environment, the programming interfaces that are provided by CodeWorld and how student programs are executed. Many of the figures illustrating this paper are created by students.  These and more can be found in the CodeWorld gallery at \url{https://code.world/gallery.html}.

To ease deployment, students need only a web browser to use CodeWorld.  They write their code with an integrated editor inside the browser. Programs are written in Haskell, which the CodeWorld server compiles to JavaScript using GHCJS~\citep{GHCJS} and sends that back to browser to execute in a canvas beside the editor. These programs are always graphical: students create static pictures, then animations, and finally interactive games and other activities.

\subsection{Two flavors of Haskell}

The standard Haskell language is not an ideal vessel for the children in CodeWorld's target audience. Therefore, CodeWorld by default provides a specially tailored educational environment. In this mode, a custom prelude is used to help students avoid common obstacles. Graphics primitives are available without an import, to create appealing visual programs. Functions of multiple arguments are not curried but rather take their arguments in a tuple, both to improve error messages and match mathematical notation that students are already learning. Finally, a single \ensuremath{\Conid{Number}} type (isomorphic to \ensuremath{\Conid{Double}}) is provided to avoid the need for type classes, and the \ensuremath{\Conid{RebindableSyntax}}\@ language extension makes literals monomorphic. Compiler error messages are post-processed to make them more intelligible to the students. Nevertheless, the code students write is still Haskell, and is accepted by GHC.

However, at \url{https://code.world/haskell} instead of \url{https://code.world/}, one finds a standard Haskell environment, with full access to the standard library. In this paper we focus on the latter variant.

\subsection{API design principles}

\begin{wrapfigure}[14]{r}{0.4\linewidth}%
\vspace{-1.1em}
\fbox{%
\adjustbox{trim={.2\width} {.2\height} {.2\width} {.2\height},scale=1.666,clip}%
{\includegraphics[width=\linewidth]{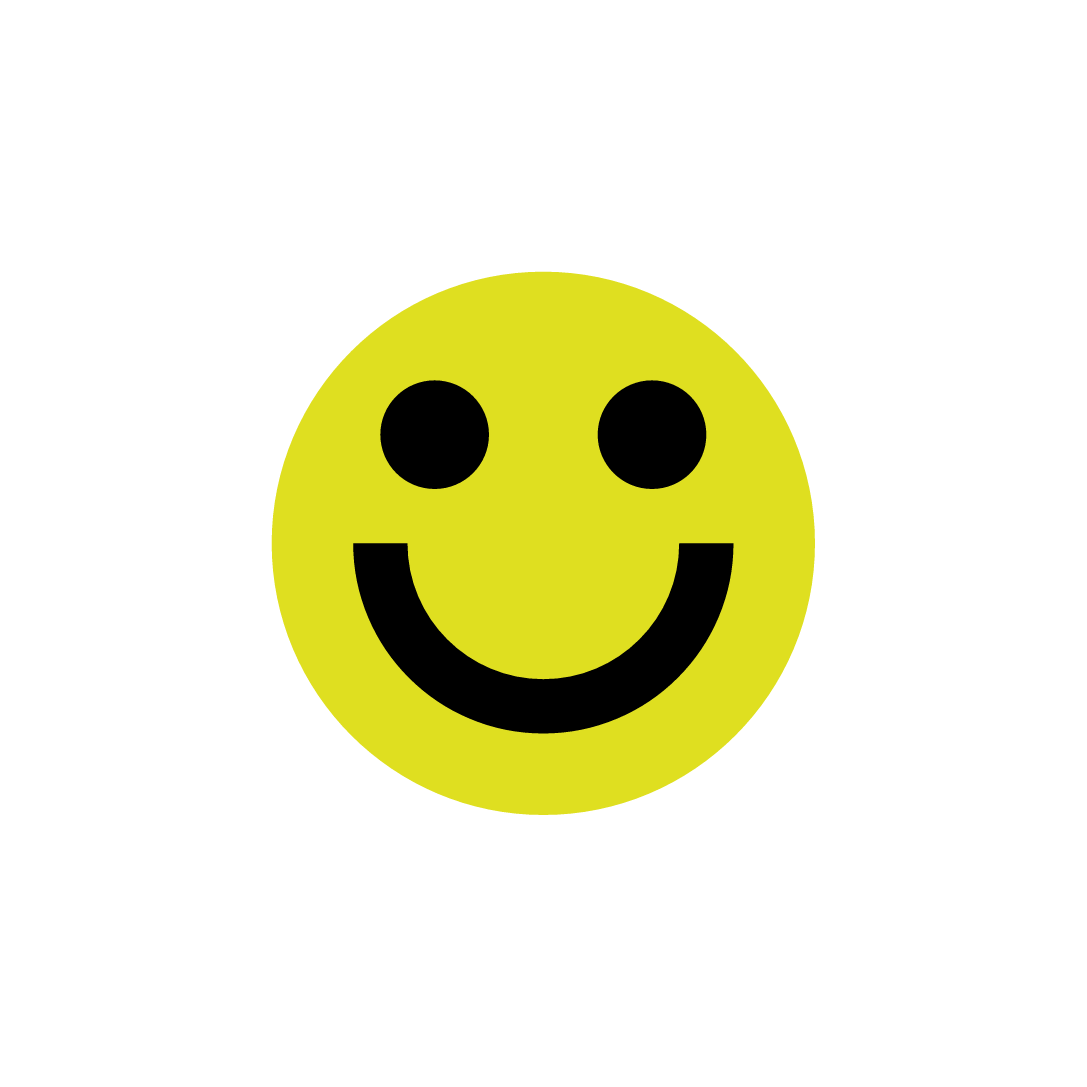}}}
\caption{Smiley}
\label{fig:smiley}
\end{wrapfigure}

An important principle of CodeWorld is to provide students with the simplest possible abstraction for a given task.  This allows them to concentrate on the ideas they want to express and think clearly about the meaning of their code, and hides as many low-level details as possible.

The first and simplest task that students face is to produce a static \emph{drawing}.  This is done with the abstract data type \ensuremath{\Conid{Picture}}, with a simple compositional API (\cref{fig:picture}) which was heavily inspired by the Gloss library \citep{gloss}.  Complex pictures are built by combining and transforming simple geometric objects.  The entry point used for this has the very simple type
\begin{hscode}\SaveRestoreHook
\column{B}{@{}>{\hspre}l<{\hspost}@{}}%
\column{E}{@{}>{\hspre}l<{\hspost}@{}}%
\>[B]{}\Varid{drawingOf}\mathbin{::}\Conid{Picture}\hsarrow{\rightarrow }{\mathpunct{.}}\Conid{IO}\;(){}\<[E]%
\ColumnHook
\end{hscode}\resethooks

\begin{figure}
\abovedisplayskip=0pt
\belowdisplayskip=0pt
\begin{hscode}\SaveRestoreHook
\column{B}{@{}>{\hspre}l<{\hspost}@{}}%
\column{17}{@{}>{\hspre}c<{\hspost}@{}}%
\column{17E}{@{}l@{}}%
\column{21}{@{}>{\hspre}l<{\hspost}@{}}%
\column{32}{@{}>{\hspre}l<{\hspost}@{}}%
\column{35}{@{}>{\hspre}l<{\hspost}@{}}%
\column{44}{@{}>{\hspre}l<{\hspost}@{}}%
\column{46}{@{}>{\hspre}l<{\hspost}@{}}%
\column{61}{@{}>{\hspre}l<{\hspost}@{}}%
\column{72}{@{}>{\hspre}l<{\hspost}@{}}%
\column{E}{@{}>{\hspre}l<{\hspost}@{}}%
\>[B]{}\textbf{\lmss data}\;\Conid{Picture}\mbox{\onelinecomment  abstract}{}\<[E]%
\\[\blanklineskip]%
\>[B]{}\mbox{\onelinecomment  Various geometric shapes (circle, rectangle, arc, polygon etc.) are}{}\<[E]%
\\[-0.3ex]%
\>[B]{}\mbox{\onelinecomment  available, and can either be filled or equipped with a thickness. E.g.:}{}\<[E]%
\\[\blanklineskip]%
\>[B]{}\mbox{\onelinecomment  Parameter: radius}{}\<[E]%
\\[-0.3ex]%
\>[B]{}\Varid{solidCircle}{}\<[17]%
\>[17]{}\mathbin{::}{}\<[17E]%
\>[21]{}\Conid{Double}\hsarrow{\rightarrow }{\mathpunct{.}}{}\<[32]%
\>[32]{}\Conid{Picture}{}\<[E]%
\\[-0.3ex]%
\>[B]{}\mbox{\onelinecomment  Parameters: thickness, radius, start angle and end angle}{}\<[E]%
\\[-0.3ex]%
\>[B]{}\Varid{thickArc}{}\<[17]%
\>[17]{}\mathbin{::}{}\<[17E]%
\>[21]{}\Conid{Double}\hsarrow{\rightarrow }{\mathpunct{.}}{}\<[35]%
\>[35]{}\Conid{Double}\hsarrow{\rightarrow }{\mathpunct{.}}{}\<[46]%
\>[46]{}\Conid{Double}\hsarrow{\rightarrow }{\mathpunct{.}}{}\<[61]%
\>[61]{}\Conid{Double}\hsarrow{\rightarrow }{\mathpunct{.}}{}\<[72]%
\>[72]{}\Conid{Picture}{}\<[E]%
\\[\blanklineskip]%
\>[B]{}\mbox{\onelinecomment  Pictures can be transformed and overlaid}{}\<[E]%
\\[-0.3ex]%
\>[B]{}\Varid{colored}{}\<[17]%
\>[17]{}\mathbin{::}{}\<[17E]%
\>[21]{}\Conid{Color}\hsarrow{\rightarrow }{\mathpunct{.}}{}\<[44]%
\>[44]{}(\Conid{Picture}\hsarrow{\rightarrow }{\mathpunct{.}}\Conid{Picture}){}\<[E]%
\\[-0.3ex]%
\>[B]{}\Varid{translated}{}\<[17]%
\>[17]{}\mathbin{::}{}\<[17E]%
\>[21]{}\Conid{Double}\hsarrow{\rightarrow }{\mathpunct{.}}\Conid{Double}\hsarrow{\rightarrow }{\mathpunct{.}}{}\<[44]%
\>[44]{}(\Conid{Picture}\hsarrow{\rightarrow }{\mathpunct{.}}\Conid{Picture}){}\<[E]%
\\[-0.3ex]%
\>[B]{}\Varid{rotated}{}\<[17]%
\>[17]{}\mathbin{::}{}\<[17E]%
\>[21]{}\Conid{Double}\hsarrow{\rightarrow }{\mathpunct{.}}{}\<[44]%
\>[44]{}(\Conid{Picture}\hsarrow{\rightarrow }{\mathpunct{.}}\Conid{Picture}){}\<[E]%
\\[-0.3ex]%
\>[B]{}\Varid{scaled}{}\<[17]%
\>[17]{}\mathbin{::}{}\<[17E]%
\>[21]{}\Conid{Double}\hsarrow{\rightarrow }{\mathpunct{.}}\Conid{Double}\hsarrow{\rightarrow }{\mathpunct{.}}{}\<[44]%
\>[44]{}(\Conid{Picture}\hsarrow{\rightarrow }{\mathpunct{.}}\Conid{Picture}){}\<[E]%
\\[-0.3ex]%
\>[B]{}(\mathbin{\&}){}\<[17]%
\>[17]{}\mathbin{::}{}\<[17E]%
\>[21]{}\Conid{Picture}\hsarrow{\rightarrow }{\mathpunct{.}}\Conid{Picture}\hsarrow{\rightarrow }{\mathpunct{.}}{}\<[44]%
\>[44]{}\Conid{Picture}{}\<[E]%
\ColumnHook
\end{hscode}\resethooks
\caption{An excerpt of CodeWorld's \ensuremath{\Conid{Picture}} API}
\label{fig:picture}
\end{figure}

This function takes care of the details of displaying the student's picture on the screen, redrawing upon window size changes and so on.  So all it takes for a student to get the computer to smile like in \cref{fig:smiley} is to write
\begin{hscode}\SaveRestoreHook
\column{B}{@{}>{\hspre}l<{\hspost}@{}}%
\column{11}{@{}>{\hspre}l<{\hspost}@{}}%
\column{E}{@{}>{\hspre}l<{\hspost}@{}}%
\>[B]{}\textbf{\lmss import}\;\Conid{CodeWorld}{}\<[E]%
\\[\blanklineskip]%
\>[B]{}\Varid{smiley}\mathrel{=}{}\<[11]%
\>[11]{}\Varid{translated}\;(\mathbin{-}\mathrm{4})\;\mathrm{4}\;(\Varid{solidCircle}\;\mathrm{2})\mathbin{\&}\Varid{translated}\;\mathrm{4}\;\mathrm{4}\;(\Varid{solidCircle}\;\mathrm{2})\mathbin{\&}{}\<[E]%
\\[-0.3ex]%
\>[11]{}\Varid{thickArc}\;\mathrm{2}\;(\mathbin{-}\Varid{pi})\;\mathrm{0}\;\mathrm{6}\mathbin{\&}\Varid{colored}\;\Varid{yellow}\;(\Varid{solidCircle}\;\mathrm{10}){}\<[E]%
\\[\blanklineskip]%
\>[B]{}\Varid{main}\mathrel{=}\Varid{drawingOf}\;\Varid{smiley}{}\<[E]%
\ColumnHook
\end{hscode}\resethooks

\begin{wrapfigure}[15]{r}{0.4\linewidth}%
\fbox{\includegraphics[width=\linewidth]{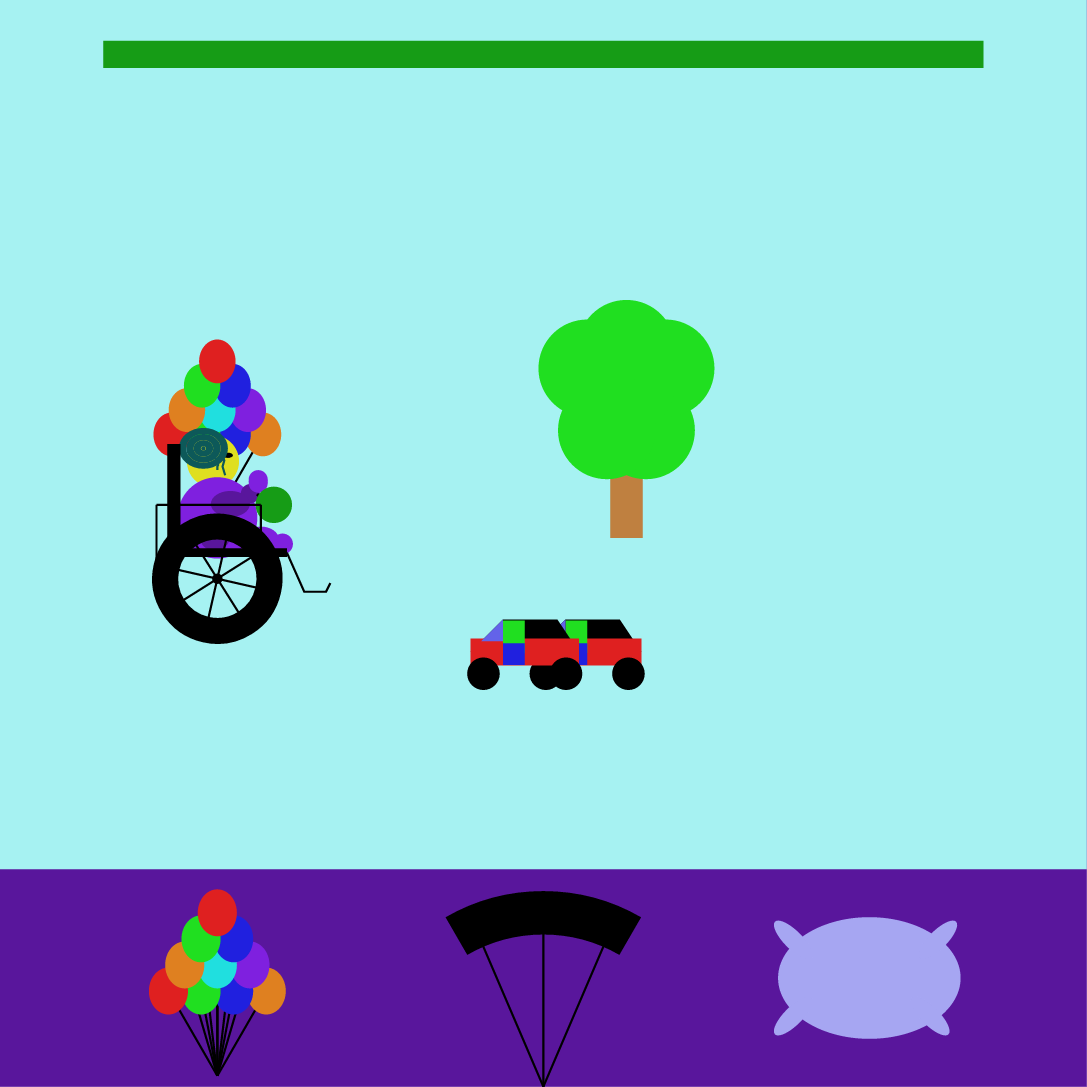}}
\caption{Yo Grandma, by Sophia (6th grade)}
\label{fig:grandma}
\end{wrapfigure}

As a next step, the students can create \emph{animations} and \emph{simulations} to make their pictures move, before eventually making their programs react to user input in \emph{interactions}. The game in \cref{fig:grandma} is a typical interaction, where the player saves flying Grandma from various obstacles by attaching balloons or parachutes to her wheelchair.

These are created by calling the following interface:
\begin{hscode}\SaveRestoreHook
\column{B}{@{}>{\hspre}l<{\hspost}@{}}%
\column{16}{@{}>{\hspre}c<{\hspost}@{}}%
\column{16E}{@{}l@{}}%
\column{20}{@{}>{\hspre}l<{\hspost}@{}}%
\column{E}{@{}>{\hspre}l<{\hspost}@{}}%
\>[B]{}\Varid{interactionOf}{}\<[16]%
\>[16]{}\mathbin{::}{}\<[16E]%
\>[20]{}\Varid{world}{}\<[E]%
\\[-0.3ex]%
\>[16]{}\hsarrow{\rightarrow }{\mathpunct{.}}{}\<[16E]%
\>[20]{}(\Conid{Double}\hsarrow{\rightarrow }{\mathpunct{.}}\Varid{world}\hsarrow{\rightarrow }{\mathpunct{.}}\Varid{world}){}\<[E]%
\\[-0.3ex]%
\>[16]{}\hsarrow{\rightarrow }{\mathpunct{.}}{}\<[16E]%
\>[20]{}(\Conid{Event}\hsarrow{\rightarrow }{\mathpunct{.}}\Varid{world}\hsarrow{\rightarrow }{\mathpunct{.}}\Varid{world}){}\<[E]%
\\[-0.3ex]%
\>[16]{}\hsarrow{\rightarrow }{\mathpunct{.}}{}\<[16E]%
\>[20]{}(\Varid{world}\hsarrow{\rightarrow }{\mathpunct{.}}\Conid{Picture}){}\<[E]%
\\[-0.3ex]%
\>[16]{}\hsarrow{\rightarrow }{\mathpunct{.}}{}\<[16E]%
\>[20]{}\Conid{IO}\;(){}\<[E]%
\ColumnHook
\end{hscode}\resethooks
In a typical call
\begin{hscode}\SaveRestoreHook
\column{B}{@{}>{\hspre}l<{\hspost}@{}}%
\column{E}{@{}>{\hspre}l<{\hspost}@{}}%
\>[B]{}\Varid{main}\mathrel{=}\Varid{interactionOf}\;\Varid{start}\;\Varid{step}\;\Varid{handle}\;\Varid{draw}{}\<[E]%
\ColumnHook
\end{hscode}\resethooks
the student passes four arguments, namely:
\begin{compactenum}
\item an initial state, \ensuremath{\Varid{start}},
\item a time step function, \ensuremath{\Varid{step}}, which calculates an updated state as time passes,
\item an event handler function, \ensuremath{\Varid{handle}}, which calculates an updated state when the user interacts with the program and
\item a visualization function, \ensuremath{\Varid{draw}}, to depict the current state as a \ensuremath{\Conid{Picture}}.
\end{compactenum}

The \ensuremath{\Conid{Event}} type, shown in \cref{fig:event}, is a simple algebraic data type that describes the press or release of a key or mouse button, or a movement of the mouse pointer.

The type of the state, \ensuremath{\Varid{world}}, is chosen by the user and consists of the domain-specific data needed by the program. The \ensuremath{\Varid{world}} type is completely unconstrained, and this will be an important factor influencing our design. It need not even be serializable, nor comparable for equality. In particular, the state may contain first-class functions and infinite lazy data structures. One way that students commonly make use of this capability is by defining infinite lazy lists of anticipated future events, based on a random number source fetched before the simulation begins.

\begin{figure}%
\abovedisplayskip=0pt
\belowdisplayskip=0pt
\begin{hscode}\SaveRestoreHook
\column{B}{@{}>{\hspre}l<{\hspost}@{}}%
\column{13}{@{}>{\hspre}c<{\hspost}@{}}%
\column{13E}{@{}l@{}}%
\column{16}{@{}>{\hspre}l<{\hspost}@{}}%
\column{E}{@{}>{\hspre}l<{\hspost}@{}}%
\>[B]{}\textbf{\lmss type}\;\Conid{Point}\mathrel{=}(\Conid{Double},\Conid{Double}){}\<[E]%
\\[-0.3ex]%
\>[B]{}\textbf{\lmss data}\;\Conid{Event}{}\<[13]%
\>[13]{}\mathrel{=}{}\<[13E]%
\>[16]{}\Conid{KeyPress}\;\Conid{Text}{}\<[E]%
\\[-0.3ex]%
\>[13]{}\mid {}\<[13E]%
\>[16]{}\Conid{KeyRelease}\;\Conid{Text}{}\<[E]%
\\[-0.3ex]%
\>[13]{}\mid {}\<[13E]%
\>[16]{}\Conid{MousePress}\;\Conid{MouseButton}\;\Conid{Point}{}\<[E]%
\\[-0.3ex]%
\>[13]{}\mid {}\<[13E]%
\>[16]{}\Conid{MouseRelease}\;\Conid{MouseButton}\;\Conid{Point}{}\<[E]%
\\[-0.3ex]%
\>[13]{}\mid {}\<[13E]%
\>[16]{}\Conid{MouseMovement}\;\Conid{Point}{}\<[E]%
\\[-0.3ex]%
\>[B]{}\textbf{\lmss data}\;\Conid{MouseButton}\mathrel{=}\Conid{LeftButton}\mid \Conid{MiddleButton}\mid \Conid{RightButton}{}\<[E]%
\ColumnHook
\end{hscode}\resethooks
\caption{The \ensuremath{\Conid{Event}} type}
\label{fig:event}
\end{figure}

\section{An interface for multi-player games}
\label{sec:api}

We would like students to extend their programming to networked multi-user programs, so that they can invite their friends to join over the internet and collaborate together on a drawing, fight each other in a fierce duel of Snake, or interact in any other way the student designs and implements. In this section, we turn our attention to choosing an API for such a task.

\subsection{Wishful thinking}
\label{sec:wishful}

Let us apply ``API design by wishful thinking", and ask: What is the most convenient abstract model of a multi-player game we can hope for, independent of implementation concerns or constraints?
\begin{wrapfigure}[16]{r}{0.4\linewidth}%
\fbox{\includegraphics[width=\linewidth]{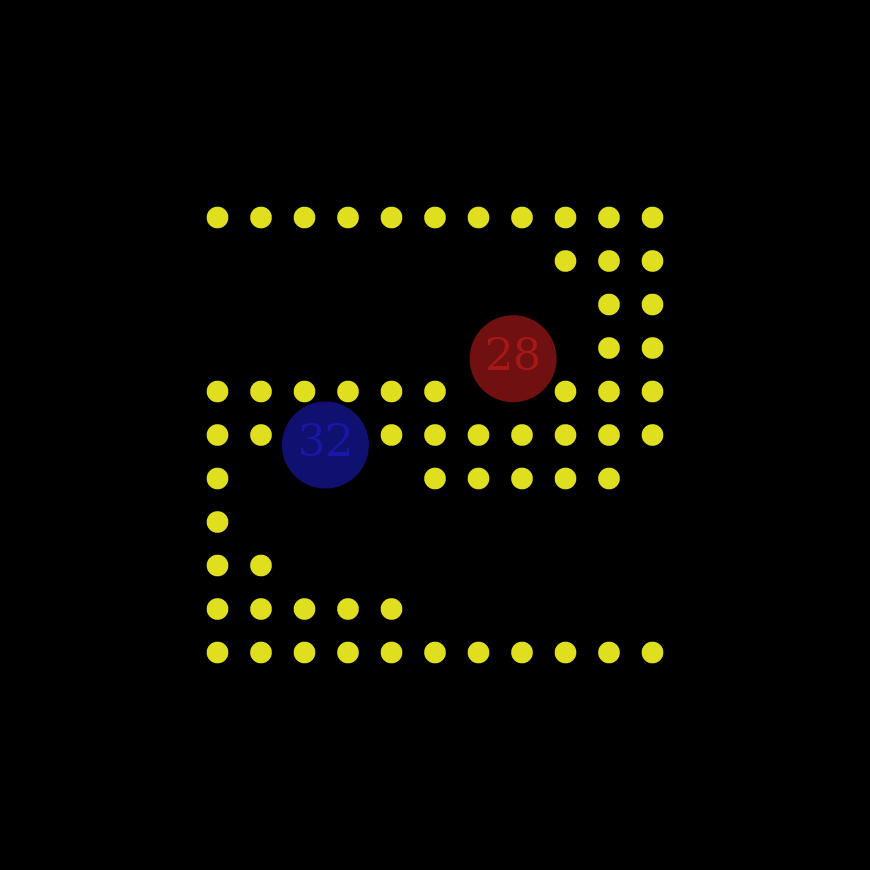}}
\caption{Dot Grab, by Adrian (7th grade)}
\label{fig:dotgrab}
\end{wrapfigure}
As experienced programmers, our thoughts might drift to network protocols or message passing between independent program instances, each with its own local state. Our students, though, care about none of this, and ideally we would not burden them with it. In fact, motivated students have already implemented games to be played with classmates, using different keys on the same device. An example is shown in \cref{fig:dotgrab}, where the red player uses the keys \keystroke{W}\keystroke{A}\keystroke{S}\keystroke{D} and the blue player the keys \UArrow\LArrow\DArrow\RArrow, in a race to consume more dots.

Their games, which they have already designed, are described in terms of one shared global state.  Why should the programming model change drastically simply because of one detail -- that the code will now run on multiple nodes communicating over a network?

We conclude, then, that an interactive multi-user program is a generalization of an interactive single-user program, and the centerpiece of the API is still a single, global state, which is mutually acted upon by all players. Basing the API on \ensuremath{\Varid{interactionOf}}, we make only minimal changes to adapt to the new environment:
\begin{itemize}
\item A new first parameter specifies the number of players.
\item The parameters \ensuremath{\Varid{start}} and \ensuremath{\Varid{step}} remain as they are.
\item The \ensuremath{\Varid{handle}} parameter, though, ought to know \emph{which} user pressed a certain button or moved their mouse, so it receives the player number (a simple \ensuremath{\Conid{Int}}) as an additional parameter.
\item Different players may also see different views of the state, so the \ensuremath{\Varid{draw}} function also receives the player number for which it should render the screen -- but it is free to ignore that parameter, of course.
\end{itemize}
All together, we arrive at the following “ideal” interface that we call \emph{collaborations}, which allows students to build networked multi-player games and other activities:
\begin{hscode}\SaveRestoreHook
\column{B}{@{}>{\hspre}l<{\hspost}@{}}%
\column{18}{@{}>{\hspre}c<{\hspost}@{}}%
\column{18E}{@{}l@{}}%
\column{22}{@{}>{\hspre}l<{\hspost}@{}}%
\column{E}{@{}>{\hspre}l<{\hspost}@{}}%
\>[B]{}\Varid{collaborationOf}{}\<[18]%
\>[18]{}\mathbin{::}{}\<[18E]%
\>[22]{}\Conid{Int}{}\<[E]%
\\[-0.3ex]%
\>[18]{}\hsarrow{\rightarrow }{\mathpunct{.}}{}\<[18E]%
\>[22]{}\Varid{world}{}\<[E]%
\\[-0.3ex]%
\>[18]{}\hsarrow{\rightarrow }{\mathpunct{.}}{}\<[18E]%
\>[22]{}(\Conid{Double}\hsarrow{\rightarrow }{\mathpunct{.}}\Varid{world}\hsarrow{\rightarrow }{\mathpunct{.}}\Varid{world}){}\<[E]%
\\[-0.3ex]%
\>[18]{}\hsarrow{\rightarrow }{\mathpunct{.}}{}\<[18E]%
\>[22]{}(\Conid{Int}\hsarrow{\rightarrow }{\mathpunct{.}}\Conid{Event}\hsarrow{\rightarrow }{\mathpunct{.}}\Varid{world}\hsarrow{\rightarrow }{\mathpunct{.}}\Varid{world}){}\<[E]%
\\[-0.3ex]%
\>[18]{}\hsarrow{\rightarrow }{\mathpunct{.}}{}\<[18E]%
\>[22]{}(\Conid{Int}\hsarrow{\rightarrow }{\mathpunct{.}}\Varid{world}\hsarrow{\rightarrow }{\mathpunct{.}}\Conid{Picture}){}\<[E]%
\\[-0.3ex]%
\>[18]{}\hsarrow{\rightarrow }{\mathpunct{.}}{}\<[18E]%
\>[22]{}\Conid{IO}\;(){}\<[E]%
\ColumnHook
\end{hscode}\resethooks

\begin{wrapfigure}[16]{r}{0.4\linewidth}%
\vspace{-2.7cm}
\fbox{\includegraphics[width=\linewidth]{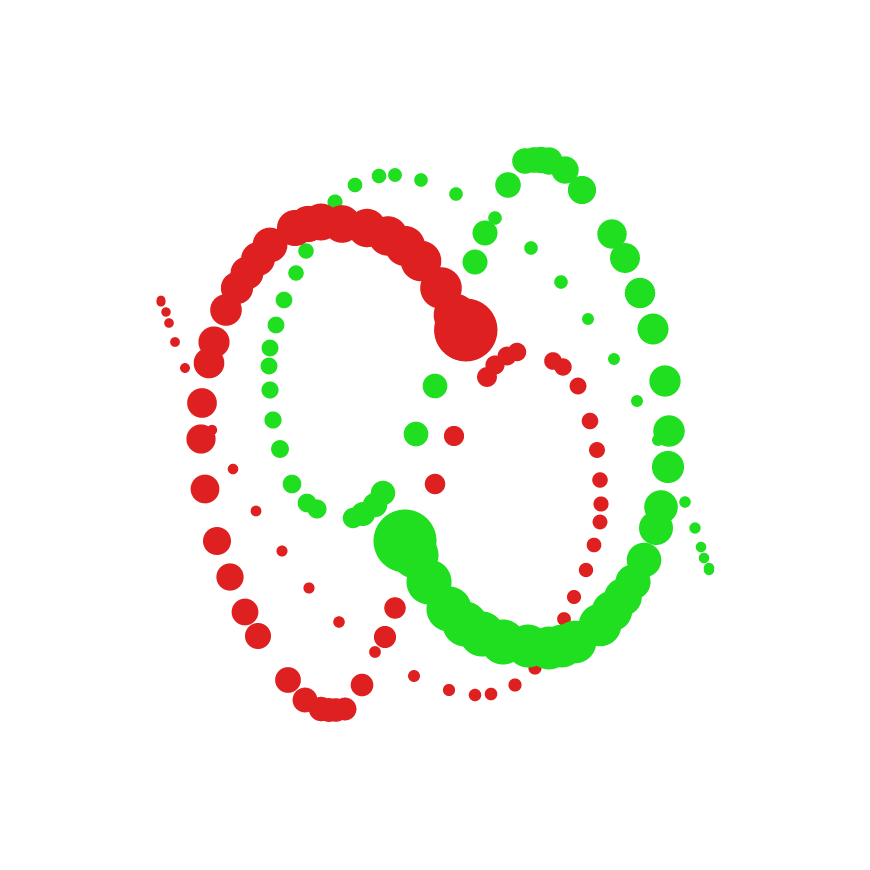}}
\caption{Two players' mouse movements}
\label{fig:traces}
\end{wrapfigure}

A small example will clarify how this interface is used. The following code traces the mouse movements of two players using colored, fading circles, and \Cref{fig:traces} shows this program in action. The green player is a bot that simply mirrors the red player's movements.

\begin{hscode}\SaveRestoreHook
\column{B}{@{}>{\hspre}l<{\hspost}@{}}%
\column{9}{@{}>{\hspre}l<{\hspost}@{}}%
\column{12}{@{}>{\hspre}l<{\hspost}@{}}%
\column{14}{@{}>{\hspre}l<{\hspost}@{}}%
\column{35}{@{}>{\hspre}l<{\hspost}@{}}%
\column{51}{@{}>{\hspre}l<{\hspost}@{}}%
\column{E}{@{}>{\hspre}l<{\hspost}@{}}%
\>[B]{}\textbf{\lmss import}\;\Conid{CodeWorld}{}\<[E]%
\\[\blanklineskip]%
\>[B]{}\textbf{\lmss type}\;\Conid{World}\mathrel{=}[\mskip1.5mu (\Conid{Color},\Conid{Double},\Conid{Double},\Conid{Double})\mskip1.5mu]{}\<[E]%
\\[\blanklineskip]%
\>[B]{}\Varid{step}\mathbin{::}\Conid{Double}\hsarrow{\rightarrow }{\mathpunct{.}}\Conid{World}\hsarrow{\rightarrow }{\mathpunct{.}}\Conid{World}{}\<[E]%
\\[-0.3ex]%
\>[B]{}\Varid{step}\;\Varid{dt}\;\Varid{dots}\mathrel{=}[\mskip1.5mu (\Varid{c},\Varid{exp}\;(\mathbin{-}\Varid{dt})\mathbin{*}\Varid{r},\Varid{x},\Varid{y})\mid (\Varid{c},\Varid{r},\Varid{x},\Varid{y})\leftarrow \Varid{dots},\Varid{r}\geq \mathrm{0.1}\mskip1.5mu]{}\<[E]%
\\[\blanklineskip]%
\>[B]{}\Varid{handle}\mathbin{::}\Conid{Int}\hsarrow{\rightarrow }{\mathpunct{.}}\Conid{Event}\hsarrow{\rightarrow }{\mathpunct{.}}\Conid{World}\hsarrow{\rightarrow }{\mathpunct{.}}\Conid{World}{}\<[E]%
\\[-0.3ex]%
\>[B]{}\Varid{handle}\;{}\<[9]%
\>[9]{}\mathrm{0}\;{}\<[12]%
\>[12]{}(\Conid{MouseMovement}\;(\Varid{x},\Varid{y}))\;{}\<[35]%
\>[35]{}\Varid{dots}\mathrel{=}(\Varid{red},{}\<[51]%
\>[51]{}\mathrm{1},\Varid{x},\Varid{y})\mathbin{:}\Varid{dots}{}\<[E]%
\\[-0.3ex]%
\>[B]{}\Varid{handle}\;{}\<[9]%
\>[9]{}\mathrm{1}\;{}\<[12]%
\>[12]{}(\Conid{MouseMovement}\;(\Varid{x},\Varid{y}))\;{}\<[35]%
\>[35]{}\Varid{dots}\mathrel{=}(\Varid{green},{}\<[51]%
\>[51]{}\mathrm{1},\Varid{x},\Varid{y})\mathbin{:}\Varid{dots}{}\<[E]%
\\[-0.3ex]%
\>[B]{}\Varid{handle}\;{}\<[9]%
\>[9]{}\anonymous \;{}\<[12]%
\>[12]{}\anonymous \;{}\<[35]%
\>[35]{}\Varid{dots}\mathrel{=}\Varid{dots}{}\<[E]%
\\[\blanklineskip]%
\>[B]{}\Varid{draw}\mathbin{::}\Conid{Int}\hsarrow{\rightarrow }{\mathpunct{.}}\Conid{World}\hsarrow{\rightarrow }{\mathpunct{.}}\Conid{Picture}{}\<[E]%
\\[-0.3ex]%
\>[B]{}\Varid{draw}\;\anonymous \;\Varid{dots}{}\<[14]%
\>[14]{}\mathrel{=}\Varid{mconcat}\;[\mskip1.5mu \Varid{translated}\;\Varid{x}\;\Varid{y}\;(\Varid{colored}\;\Varid{c}\;(\Varid{solidCircle}\;\Varid{r}))\mid (\Varid{c},\Varid{r},\Varid{x},\Varid{y})\leftarrow \Varid{dots}\mskip1.5mu]{}\<[E]%
\\[\blanklineskip]%
\>[B]{}\Varid{main}\mathbin{::}\Conid{IO}\;(){}\<[E]%
\\[-0.3ex]%
\>[B]{}\Varid{main}\mathrel{=}\Varid{collaborationOf}\;\mathrm{2}\;[\mskip1.5mu \mskip1.5mu]\;\Varid{step}\;\Varid{handle}\;\Varid{draw}{}\<[E]%
\ColumnHook
\end{hscode}\resethooks
\goodbreak

A collaboration begins with a lobby, featuring buttons to create or join a game. Upon creating a new game, the player is given a four-letter code to be shared with friends. Those friends may enter the four-letter code to join the game. Once enough players have joined, the game begins.

\subsection{Solving random problems with the module system}
\label{sec:no-io}

Like \ensuremath{\Varid{interactionOf}} before it, the parameters of \ensuremath{\Varid{collaborationOf}} provide enough information to completely determine the behavior of the program from the sequence of time steps and UI events that occur. Unlike \ensuremath{\Varid{interactionOf}}, however, a collaboration involves \emph{more than one} use of the \ensuremath{\Varid{collaborationOf}} API, as the function is executed by each participating player. To ensure that there is a single, well-defined behavior, it is essential that all players run \ensuremath{\Varid{collaborationOf}} with the same arguments. Obviously, we need to ensure that all clients run the same program, and the CodeWorld server does so. But even with the same code, the arguments to \ensuremath{\Varid{collaborationOf}} can differ from client to client:
\begin{hscode}\SaveRestoreHook
\column{B}{@{}>{\hspre}l<{\hspost}@{}}%
\column{3}{@{}>{\hspre}l<{\hspost}@{}}%
\column{E}{@{}>{\hspre}l<{\hspost}@{}}%
\>[B]{}\Varid{main}\mathrel{=}\textbf{\lmss do}{}\<[E]%
\\[-0.3ex]%
\>[B]{}\hsindent{3}{}\<[3]%
\>[3]{}\Varid{r}\leftarrow \Varid{randomRIO}\;(\mathrm{0},\mathrm{1}){}\<[E]%
\\[-0.3ex]%
\>[B]{}\hsindent{3}{}\<[3]%
\>[3]{}\Varid{collaborationOf}\;\Varid{numPlayers}\;\Varid{start}\;\Varid{step}\;(\Varid{handle}\;\Varid{r})\;\Varid{draw}{}\<[E]%
\ColumnHook
\end{hscode}\resethooks
The event handling function now depends on I/O -- specifically, the choice of a random number -- and it is very unlikely that all clients happen to pick the same random number. Despite sharing the same code, the clients will disagree about the correct behavior of the system.

The problem is not the use of random numbers per se, but rather the unconstrained flow of client-specific state resulting from \emph{any} I/O into the \ensuremath{\Varid{collaborationOf}} API via free variables in its parameters. Since most of the parameters to \ensuremath{\Varid{collaborationOf}} have function types, we cannot just compare them to establish consistency at runtime.

We solve this problem in two ways: one in the educational environment, and the other in the standard Haskell environment.

In the former, we have tight control over the set of library functions available to the student. No packages are exposed except for a custom standard library with a heavily customized \ensuremath{\Conid{Prelude}} module, and this library simply does not provide any functions to compose IO operations, such as as the monadic bind operators (\ensuremath{\bind }, \ensuremath{\sequ }). This also rules out the use of Haskell's \ensuremath{\textbf{\lmss do}}-notation, which under the regime of \ensuremath{\Conid{RebindableSyntax}} requires an operator called (\ensuremath{\bind }) to be in scope. A valid Haskell program requires a top-level function \ensuremath{\Varid{main}\mathbin{::}\Conid{IO}\;()}, and since the only available way to obtain an \ensuremath{\Conid{IO}\;()} is through our API entry points (\ensuremath{\Varid{drawingOf}}, \ensuremath{\Varid{interactionOf}}, and so on), we know that all CodeWorld collaborations are of essentially the form \ensuremath{\Varid{main}\mathrel{=}\Varid{collaborationOf}\ldots} In particular, no I/O can be executed prior to the collaboration, and hence no client-dependent behavior is possible.

\subsection{Solving random problems syntactically}
\label{sec:static-pointers}

This solution is not suitable for the standard Haskell environment, where we do not want to restrict the user's access to the standard library. We can still prevent the user from using the results of client-specific I/O in arguments to \ensuremath{\Varid{collaborationOf}}.  To accomplish this, we creatively use the work of \citet{static-pointers}, who sought to bring Erlang-like distributed computing to Haskell. They had to exchange functions over the network, which is possible by passing code references, as long as no potentially unserializable values are captured from the environment.  To guarantee that, they introduced a Haskell language extension, \emph{static pointers}, which introduces:
\begin{compactitem}
\item a new type constructor \ensuremath{\Conid{StaticPtr}\;\Varid{a}}, which wraps values of type \ensuremath{\Varid{a}},
\item a new syntactic construct \ensuremath{{\textbf{\lmss static}}\;\Varid{foo}}, such that for any expression \ensuremath{\Varid{foo}} of type \ensuremath{\Varid{a}}, the expression \ensuremath{{\textbf{\lmss static}}\;\Varid{foo}} has type \ensuremath{\Conid{StaticPtr}\;\Varid{a}}, but is only valid if \ensuremath{\Varid{foo}} does not contain any locally bound free variables,
\item a pure function \ensuremath{\Varid{deRefStaticPtr}\mathbin{::}\Conid{StaticPtr}\;\Varid{a}\hsarrow{\rightarrow }{\mathpunct{.}}\Varid{a}}, to unwrap the static pointer, and
\item a pure function \ensuremath{\Varid{staticKey}\mathbin{::}\Conid{StaticPtr}\;\Varid{a}\hsarrow{\rightarrow }{\mathpunct{.}}\Conid{StaticKey}} which produces a key that -- within one program -- uniquely identifies a static pointer.
\end{compactitem}
The requirement that \ensuremath{\Conid{StaticPtr}} values cannot have locally bound free variables turns out to be exactly what we need to prevent programs from smuggling client-specific state obtained with I/O actions into collaborations. We therefore further refine the API to require its arguments to be static pointers:
\begin{hscode}\SaveRestoreHook
\column{B}{@{}>{\hspre}l<{\hspost}@{}}%
\column{18}{@{}>{\hspre}c<{\hspost}@{}}%
\column{18E}{@{}l@{}}%
\column{22}{@{}>{\hspre}l<{\hspost}@{}}%
\column{E}{@{}>{\hspre}l<{\hspost}@{}}%
\>[B]{}\Varid{collaborationOf}{}\<[18]%
\>[18]{}\mathbin{::}{}\<[18E]%
\>[22]{}\Conid{Int}{}\<[E]%
\\[-0.3ex]%
\>[18]{}\hsarrow{\rightarrow }{\mathpunct{.}}{}\<[18E]%
\>[22]{}\Conid{StaticPtr}\;\Varid{world}{}\<[E]%
\\[-0.3ex]%
\>[18]{}\hsarrow{\rightarrow }{\mathpunct{.}}{}\<[18E]%
\>[22]{}\Conid{StaticPtr}\;(\Conid{Double}\hsarrow{\rightarrow }{\mathpunct{.}}\Varid{world}\hsarrow{\rightarrow }{\mathpunct{.}}\Varid{world}){}\<[E]%
\\[-0.3ex]%
\>[18]{}\hsarrow{\rightarrow }{\mathpunct{.}}{}\<[18E]%
\>[22]{}\Conid{StaticPtr}\;(\Conid{Int}\hsarrow{\rightarrow }{\mathpunct{.}}\Conid{Event}\hsarrow{\rightarrow }{\mathpunct{.}}\Varid{world}\hsarrow{\rightarrow }{\mathpunct{.}}\Varid{world}){}\<[E]%
\\[-0.3ex]%
\>[18]{}\hsarrow{\rightarrow }{\mathpunct{.}}{}\<[18E]%
\>[22]{}\Conid{StaticPtr}\;(\Conid{Int}\hsarrow{\rightarrow }{\mathpunct{.}}\Varid{world}\hsarrow{\rightarrow }{\mathpunct{.}}\Conid{Picture}){}\<[E]%
\\[-0.3ex]%
\>[18]{}\hsarrow{\rightarrow }{\mathpunct{.}}{}\<[18E]%
\>[22]{}\Conid{IO}\;(){}\<[E]%
\ColumnHook
\end{hscode}\resethooks
The mouse tracing program in \Cref{fig:traces} must now change its definition of \ensuremath{\Varid{main}} to
\begin{hscode}\SaveRestoreHook
\column{B}{@{}>{\hspre}l<{\hspost}@{}}%
\column{E}{@{}>{\hspre}l<{\hspost}@{}}%
\>[B]{}\Varid{main}\mathrel{=}\Varid{collaborationOf}\;\mathrm{2}\;({\textbf{\lmss static}}\;[\mskip1.5mu \mskip1.5mu])\;({\textbf{\lmss static}}\;\Varid{step})\;({\textbf{\lmss static}}\;\Varid{handle})\;({\textbf{\lmss static}}\;\Varid{draw}).{}\<[E]%
\ColumnHook
\end{hscode}\resethooks
On the other hand, writing \ensuremath{{\textbf{\lmss static}}\;(\Varid{handle}\;\Varid{r})} to smuggle in a randomly drawn number \ensuremath{\Varid{r}}, as in the example above, will fail at compile time.
\iflong
Requiring the \ensuremath{{\textbf{\lmss static}}} keyword here admittedly muddies the clarity of the API a bit. We believe that the target audience of CodeWorld's standard Haskell mode can handle this. Beginners working within the educational mode need not deal with this slight complication.

\fi
A somewhat more clever attempt, though, still causes problems:
\goodbreak
\begin{hscode}\SaveRestoreHook
\column{B}{@{}>{\hspre}l<{\hspost}@{}}%
\column{3}{@{}>{\hspre}l<{\hspost}@{}}%
\column{E}{@{}>{\hspre}l<{\hspost}@{}}%
\>[B]{}\Varid{main}\mathrel{=}\textbf{\lmss do}{}\<[E]%
\\[-0.3ex]%
\>[B]{}\hsindent{3}{}\<[3]%
\>[3]{}\Varid{coinFlip}\leftarrow \Varid{randomIO}{}\<[E]%
\\[-0.3ex]%
\>[B]{}\hsindent{3}{}\<[3]%
\>[3]{}\textbf{\lmss let}\;\Varid{step}\mathrel{=}\textbf{\lmss if}\;\Varid{coinFlip}\;\textbf{\lmss then}\;{\textbf{\lmss static}}\;\Varid{step1}\;\textbf{\lmss else}\;{\textbf{\lmss static}}\;\Varid{step2}{}\<[E]%
\\[-0.3ex]%
\>[B]{}\hsindent{3}{}\<[3]%
\>[3]{}\Varid{collaborationOf}\;\mathrm{2}\;({\textbf{\lmss static}}\;[\mskip1.5mu \mskip1.5mu])\;\Varid{step}\;({\textbf{\lmss static}}\;\Varid{handle})\;({\textbf{\lmss static}}\;\Varid{draw}){}\<[E]%
\ColumnHook
\end{hscode}\resethooks
This program is accepted by the compiler because the arguments to \ensuremath{\Varid{collaborationOf}} are indeed \ensuremath{\Conid{StaticPtr}} values of the right types, yet it raises the same questions when clients disagree on the choice of step function. While we cannot prevent this case at compile time, we can at least detect it at runtime. Static pointers can be serialized using the function \ensuremath{\Varid{staticKey}\mathbin{::}\Conid{StaticPtr}\;\Varid{a}\hsarrow{\rightarrow }{\mathpunct{.}}\Conid{StaticKey}}. Before a game starts, the participating clients compare the keys of their arguments to check that they match. This is a subtly different use of static pointers from the original intent of sending functions over a network in a message-passing protocol. We need not actually receive the original values on the remote end of our connections, but instead use the serialized keys only to check for consistency.

With this check in place -- short of using unsafe features such as \ensuremath{\Varid{unsafePerformIO}} -- we are confident that every client is indeed running the same functions. However, this forces our games to be entirely deterministic. This is a problem, since many games involve an element of chance! To restore the possibility of random behavior, we supply a random number source to use in building the initial state, with a consistent seed in all clients. The type of the \ensuremath{\Varid{start}} parameter is now \ensuremath{\Conid{StaticPtr}\;(\Conid{StdGen}\hsarrow{\rightarrow }{\mathpunct{.}}\Varid{world})}.
\iflong
(This is not entirely new: CodeWorld's educational environment has never exported a random number generator, and its simulations and interactions have always been initialized with an infinite list of random numbers.)
\fi

This completes our derivation of \ensuremath{\Varid{collaborationOf}}, which in its final form is
\begin{hscode}\SaveRestoreHook
\column{B}{@{}>{\hspre}l<{\hspost}@{}}%
\column{18}{@{}>{\hspre}c<{\hspost}@{}}%
\column{18E}{@{}l@{}}%
\column{22}{@{}>{\hspre}l<{\hspost}@{}}%
\column{E}{@{}>{\hspre}l<{\hspost}@{}}%
\>[B]{}\Varid{collaborationOf}{}\<[18]%
\>[18]{}\mathbin{::}{}\<[18E]%
\>[22]{}\Conid{Int}{}\<[E]%
\\[-0.3ex]%
\>[18]{}\hsarrow{\rightarrow }{\mathpunct{.}}{}\<[18E]%
\>[22]{}\Conid{StaticPtr}\;(\Conid{StdGen}\hsarrow{\rightarrow }{\mathpunct{.}}\Varid{world}){}\<[E]%
\\[-0.3ex]%
\>[18]{}\hsarrow{\rightarrow }{\mathpunct{.}}{}\<[18E]%
\>[22]{}\Conid{StaticPtr}\;(\Conid{Double}\hsarrow{\rightarrow }{\mathpunct{.}}\Varid{world}\hsarrow{\rightarrow }{\mathpunct{.}}\Varid{world}){}\<[E]%
\\[-0.3ex]%
\>[18]{}\hsarrow{\rightarrow }{\mathpunct{.}}{}\<[18E]%
\>[22]{}\Conid{StaticPtr}\;(\Conid{Int}\hsarrow{\rightarrow }{\mathpunct{.}}\Conid{Event}\hsarrow{\rightarrow }{\mathpunct{.}}\Varid{world}\hsarrow{\rightarrow }{\mathpunct{.}}\Varid{world}){}\<[E]%
\\[-0.3ex]%
\>[18]{}\hsarrow{\rightarrow }{\mathpunct{.}}{}\<[18E]%
\>[22]{}\Conid{StaticPtr}\;(\Conid{Int}\hsarrow{\rightarrow }{\mathpunct{.}}\Varid{world}\hsarrow{\rightarrow }{\mathpunct{.}}\Conid{Picture}){}\<[E]%
\\[-0.3ex]%
\>[18]{}\hsarrow{\rightarrow }{\mathpunct{.}}{}\<[18E]%
\>[22]{}\Conid{IO}\;(){}\<[E]%
\ColumnHook
\end{hscode}\resethooks

\section{From wishful thinking to running code}
\label{sec:prediction}

How can we implement this interface? It turns out that our implementation options are severely narrowed down by the following requirements:

\begin{compactenum}
\item We need to handle any code using the API. Given the educational setting of CodeWorld, we cannot require any particular discipline.
\item The players need to see an eventually consistent state. They may have different ideas about the state of the world, but only until everybody receives information about everybody’s interactions.
\item The effects of a player's own interactions are immediately visible to that player.
Even a “local” interaction, such as selecting a piece in a game of Chess, will have to represented in the game state, and any latency here would make the user interface sluggish.
\end{compactenum}

The first requirement in particular implies that the game state is completely opaque to us. This already rules out the usual client-server architecture, where only the central server manages the game state and the clients send abstract moves (e.g., ``white moves the knight to e8") and render the game state that they receive from the server. We have neither insight into what constitutes an abstract move, nor how to serialize and transmit the game state.

We could avoid this problem by sending the raw UI \ensuremath{\Conid{Event}} instead of an abstract move to the server, and letting the server respond to each client with the \ensuremath{\Conid{Picture}} to show. This ``dumb terminal'' approach however would run afoul of our third requirement, as every user interaction would be delayed by the time it takes messages to travel to the server and back.

The requirement of immediate responsiveness implies that every client needs to manage its own copy of the game state, and being abstract in the game state implies that there is nothing else but the UI events that the clients can transmit to synchronize the state. In other words, lock-step simulation is the only way for us.

\iflong
This approach assumes the integrity of client code. Since all clients track the entire game state, malicious players could trick CodeWorld into running a modified version of the program which, among other things, could then reveal hidden parts of the game state. Given the educational goals of CodeWorld, we are willing to trade this security for a cleaner API.
\fi

\subsection{Types and messages}

We seek, then, to implement the API by exchanging UI events between clients. For the purposes of this paper, it does not matter how events are transmitted from client to client. The CodeWorld implementation uses a very simple relay server that broadcasts messages from one client to the others via WebSockets (a full-duplex server-client protocol for web applications), but peer-to-peer communication using WebRTC (a peer-to-peer protocol for web applications) or other methods would work equally well, as long as they deliver events reliably and in order.

Every such message obviously needs to contain the actual \ensuremath{\Conid{Event}} and the player number. In addition, it must contain a timestamp, so that each client applies the event at the same time despite differences in network latency. Otherwise -- assuming a time-sensitive game with a non-trivial \ensuremath{\Varid{step}} function -- the various clients would obtain different views of the world. Timestamps are \ensuremath{\Conid{Double}} values, measured in seconds since the start of the game.

\begin{hscode}\SaveRestoreHook
\column{B}{@{}>{\hspre}l<{\hspost}@{}}%
\column{17}{@{}>{\hspre}c<{\hspost}@{}}%
\column{17E}{@{}l@{}}%
\column{20}{@{}>{\hspre}l<{\hspost}@{}}%
\column{E}{@{}>{\hspre}l<{\hspost}@{}}%
\>[B]{}\textbf{\lmss type}\;\Conid{Timestamp}{}\<[17]%
\>[17]{}\mathrel{=}{}\<[17E]%
\>[20]{}\Conid{Double}{}\<[E]%
\\[-0.3ex]%
\>[B]{}\textbf{\lmss type}\;\Conid{Player}{}\<[17]%
\>[17]{}\mathrel{=}{}\<[17E]%
\>[20]{}\Conid{Int}{}\<[E]%
\\[-0.3ex]%
\>[B]{}\textbf{\lmss type}\;\Conid{Message}{}\<[17]%
\>[17]{}\mathrel{=}{}\<[17E]%
\>[20]{}(\Conid{Timestamp},\Conid{Player},\Conid{Event}){}\<[E]%
\ColumnHook
\end{hscode}\resethooks

\subsection{Resettable state}

Having fixed the message type still leaves open the question of what to do with these messages, which is non-trivial due to the network latency.

Assume that 23.5 seconds into a real-time strategy game, I send my knights to attack the other player. My client sends the corresponding message \ensuremath{(\mathrm{23.500},\mathrm{0},\Conid{MousePress}\;\Conid{LeftButton}\;(\mathrm{20},\mathrm{30}))} to the other player. The message arrives, say, 100ms later. As mentioned before, the other player cannot simply let my knights set out a bit later. What else?

The classical solution \citep{terranobettner} is to not act on local events immediately, but add a delay of, say, 200ms. The message would be \ensuremath{(\mathrm{23.700},\mathrm{0},\Conid{MousePress}\;\Conid{LeftButton}\;(\mathrm{20},\mathrm{30}))}, and assuming it reaches all other players in time, all are able to apply the event at precisely the same moment. This solution works well if the UI can somehow respond to the user's actions immediately, e.g.\@ by letting the knight audibly confirm the command, so hide this delay from the user.

The luxury of such a separation is not available to us -- according to the third requirement, each client must immediately apply its own events -- and the message really has to have the timestamp 23.500. This leaves the other player, when it receives the message 100ms later, with no choice but to roll back the game state to time 23.500, apply my event, and replay the following 100ms. While rollback and replay are hard to implement in imperative programming paradigms, where every piece of data can have local mutable state, they are easy in Haskell, where we know that the value of type \ensuremath{\Varid{world}} really holds all relevant bits of the program's state.

One way of allowing such recalculation is to simply not store the state at all, and re-calculate it every time we draw the game screen. The function to do so would expect the game specification, the current time and the list of messages that we have seen so far, including the locally generated ones, and would calculate the game state. Its type signature would thus be
\begin{hscode}\SaveRestoreHook
\column{B}{@{}>{\hspre}l<{\hspost}@{}}%
\column{E}{@{}>{\hspre}l<{\hspost}@{}}%
\>[B]{}\Varid{currentState}\mathbin{::}\Conid{Game}\;\Varid{world}\Rightarrow \Conid{Timestamp}\hsarrow{\rightarrow }{\mathpunct{.}}[\mskip1.5mu \Conid{Message}\mskip1.5mu]\hsarrow{\rightarrow }{\mathpunct{.}}\Varid{world}{}\<[E]%
\ColumnHook
\end{hscode}\resethooks
where the hypothetical type class \ensuremath{\Conid{Game}} captures the user-defined game logic; we introduce it here to avoid obscuring the following code listings by passing it explicitly around as an argument:
\begin{hscode}\SaveRestoreHook
\column{B}{@{}>{\hspre}l<{\hspost}@{}}%
\column{3}{@{}>{\hspre}l<{\hspost}@{}}%
\column{11}{@{}>{\hspre}c<{\hspost}@{}}%
\column{11E}{@{}l@{}}%
\column{15}{@{}>{\hspre}l<{\hspost}@{}}%
\column{E}{@{}>{\hspre}l<{\hspost}@{}}%
\>[B]{}\textbf{\lmss class}\;\Conid{Game}\;\Varid{world}\;\textbf{\lmss where}{}\<[E]%
\\[-0.3ex]%
\>[B]{}\hsindent{3}{}\<[3]%
\>[3]{}\Varid{start}{}\<[11]%
\>[11]{}\mathbin{::}{}\<[11E]%
\>[15]{}\Varid{world}{}\<[E]%
\\[-0.3ex]%
\>[B]{}\hsindent{3}{}\<[3]%
\>[3]{}\Varid{step}{}\<[11]%
\>[11]{}\mathbin{::}{}\<[11E]%
\>[15]{}\Conid{Double}\hsarrow{\rightarrow }{\mathpunct{.}}\Varid{world}\hsarrow{\rightarrow }{\mathpunct{.}}\Varid{world}{}\<[E]%
\\[-0.3ex]%
\>[B]{}\hsindent{3}{}\<[3]%
\>[3]{}\Varid{handle}{}\<[11]%
\>[11]{}\mathbin{::}{}\<[11E]%
\>[15]{}\Conid{Player}\hsarrow{\rightarrow }{\mathpunct{.}}\Conid{Event}\hsarrow{\rightarrow }{\mathpunct{.}}\Varid{world}\hsarrow{\rightarrow }{\mathpunct{.}}\Varid{world}{}\<[E]%
\ColumnHook
\end{hscode}\resethooks

Assume, for a short while, that there was no \ensuremath{\Varid{step}} function, i.e.\@ the game state changes only when there is an actual event. Then the timestamps are only required to put the events into the right order and to disregard events which are not yet to be applied (which can happen if the player’s game time started at slightly different points in time):
\begin{hscode}\SaveRestoreHook
\column{B}{@{}>{\hspre}l<{\hspost}@{}}%
\column{3}{@{}>{\hspre}l<{\hspost}@{}}%
\column{10}{@{}>{\hspre}l<{\hspost}@{}}%
\column{E}{@{}>{\hspre}l<{\hspost}@{}}%
\>[B]{}\Varid{currentState}\mathbin{::}\Conid{Game}\;\Varid{world}\Rightarrow \Conid{Timestamp}\hsarrow{\rightarrow }{\mathpunct{.}}[\mskip1.5mu \Conid{Message}\mskip1.5mu]\hsarrow{\rightarrow }{\mathpunct{.}}\Varid{world}{}\<[E]%
\\[-0.3ex]%
\>[B]{}\Varid{currentState}\;\Varid{now}\;\Varid{messages}\mathrel{=}\Varid{applyEvents}\;\Varid{to\char95 apply}\;\Varid{start}{}\<[E]%
\\[-0.3ex]%
\>[B]{}\hsindent{3}{}\<[3]%
\>[3]{}\textbf{\lmss where}\;\Varid{to\char95 apply}\mathrel{=}\Varid{takeWhile}\;(\mathbf{\lmss\uplambda}\hslambda (\Varid{t},\anonymous ,\anonymous )\hsarrow{\rightarrow }{\mathpunct{.}}\Varid{t}\leq \Varid{now})\;(\Varid{sortMessages}\;\Varid{messages}){}\<[E]%
\\[\blanklineskip]%
\>[B]{}\Varid{sortMessages}\mathbin{::}[\mskip1.5mu \Conid{Messages}\mskip1.5mu]\hsarrow{\rightarrow }{\mathpunct{.}}[\mskip1.5mu \Conid{Messages}\mskip1.5mu]{}\<[E]%
\\[-0.3ex]%
\>[B]{}\Varid{sortMessages}\mathrel{=}\Varid{sortOn}\;(\mathbf{\lmss\uplambda}\hslambda (\Varid{t},\Varid{p},\anonymous )\hsarrow{\rightarrow }{\mathpunct{.}}(\Varid{t},\Varid{p}))\;\Varid{messages}{}\<[E]%
\\[\blanklineskip]%
\>[B]{}\Varid{applyEvents}\mathbin{::}\Conid{Game}\;\Varid{world}\Rightarrow [\mskip1.5mu \Conid{Message}\mskip1.5mu]\hsarrow{\rightarrow }{\mathpunct{.}}\Varid{world}\hsarrow{\rightarrow }{\mathpunct{.}}\Varid{world}{}\<[E]%
\\[-0.3ex]%
\>[B]{}\Varid{applyEvents}\;\Varid{messages}\;\Varid{w}\mathrel{=}\Varid{foldl}\;\Varid{apply}\;\Varid{w}\;\Varid{messages}{}\<[E]%
\\[-0.3ex]%
\>[B]{}\hsindent{3}{}\<[3]%
\>[3]{}\textbf{\lmss where}\;{}\<[10]%
\>[10]{}\Varid{apply}\;\Varid{w}\;(\anonymous ,\Varid{p},\Varid{e})\mathrel{=}\Varid{handle}\;\Varid{p}\;\Varid{e}\;\Varid{w}{}\<[E]%
\ColumnHook
\end{hscode}\resethooks

Eventually, every client receives the same list of messages, up to the interleaving of events from different players. After a stable sort by timestamp and player, the lists of events will be identical, so all clients will calculate the same game state.

\subsection{A few more steps}

This is nice and simple, but ignores the \ensuremath{\Varid{step}} function, which models the evolution of the state as time passes. Clearly, we have to call \ensuremath{\Varid{step}} before each event, and again at the end. In order to calculate the time passed since the last event, we also have to keep track of which timestamp a snapshot of the game state corresponds to:

\begin{hscode}\SaveRestoreHook
\column{B}{@{}>{\hspre}l<{\hspost}@{}}%
\column{3}{@{}>{\hspre}l<{\hspost}@{}}%
\column{10}{@{}>{\hspre}l<{\hspost}@{}}%
\column{22}{@{}>{\hspre}l<{\hspost}@{}}%
\column{E}{@{}>{\hspre}l<{\hspost}@{}}%
\>[B]{}\Varid{currentState}\mathbin{::}\Conid{Game}\;\Varid{world}\Rightarrow \Conid{Timestamp}\hsarrow{\rightarrow }{\mathpunct{.}}[\mskip1.5mu \Conid{Message}\mskip1.5mu]\hsarrow{\rightarrow }{\mathpunct{.}}\Varid{world}{}\<[E]%
\\[-0.3ex]%
\>[B]{}\Varid{currentState}\;\Varid{now}\;\Varid{messages}\mathrel{=}\Varid{step}\;(\Varid{now}\mathbin{-}\Varid{t})\;\Varid{world}{}\<[E]%
\\[-0.3ex]%
\>[B]{}\hsindent{3}{}\<[3]%
\>[3]{}\textbf{\lmss where}\;{}\<[10]%
\>[10]{}\Varid{to\char95 apply}{}\<[22]%
\>[22]{}\mathrel{=}\Varid{takeWhile}\;(\mathbf{\lmss\uplambda}\hslambda (\Varid{t},\anonymous ,\anonymous )\hsarrow{\rightarrow }{\mathpunct{.}}\Varid{t}\leq \Varid{now})\;(\Varid{sortMessages}\;\Varid{messages}){}\<[E]%
\\[-0.3ex]%
\>[10]{}(\Varid{t},\Varid{world}){}\<[22]%
\>[22]{}\mathrel{=}\Varid{applyEvents}\;\Varid{to\char95 apply}\;(\mathrm{0},\Varid{start}){}\<[E]%
\\[\blanklineskip]%
\>[B]{}\Varid{applyEvents}\mathbin{::}\Conid{Game}\;\Varid{world}\Rightarrow [\mskip1.5mu \Conid{Message}\mskip1.5mu]\hsarrow{\rightarrow }{\mathpunct{.}}(\Conid{Timestamp},\Varid{world})\hsarrow{\rightarrow }{\mathpunct{.}}(\Conid{Timestamp},\Varid{world}){}\<[E]%
\\[-0.3ex]%
\>[B]{}\Varid{applyEvents}\;\Varid{messages}\;\Varid{ts}\mathrel{=}\Varid{foldl}\;\Varid{apply}\;\Varid{ts}\;\Varid{messages}{}\<[E]%
\\[-0.3ex]%
\>[B]{}\hsindent{3}{}\<[3]%
\>[3]{}\textbf{\lmss where}\;{}\<[10]%
\>[10]{}\Varid{apply}\;(\Varid{t0},\Varid{world})\;(\Varid{t1},\Varid{p},\Varid{e})\mathrel{=}(\Varid{t1},\Varid{handle}\;\Varid{p}\;\Varid{e}\;(\Varid{step}\;(\Varid{t1}\mathbin{-}\Varid{t0})\;\Varid{world}))){}\<[E]%
\ColumnHook
\end{hscode}\resethooks

Unfortunately, students would not be quite happy with this implementation. The \ensuremath{\Varid{step}} function is commonly used to calculate a single step in a physics simulation, which requires that it is called often enough to achieve a decent simulation frequency.

For instance, when simulating a projectile, a common technique is to adjust the position linearly along the velocity vector, and the velocity linearly according to forces like gravity or drag. The result is a stepwise-linear approximation, the precision of which depends on the sampling frequency. Another common technique is to do collision detection only once per time step, and again the result depends on the frequency of steps.  It is important, then, that the \ensuremath{\Varid{step}} function is called at a reasonably high frequency.

We could leave students to resolve this themselves, by dividing time steps into multiple finer steps, if necessary, in their \ensuremath{\Varid{step}} implementation.  However, imposing that burden would violate our first requirement: not requiring any discipline from the user. Therefore, we have to ensure that the \ensuremath{\Varid{step}} function is called often enough, even if there is no user event for a while.

In simulations and interactions, the implemented behavior is to evaluate the step function as quickly as possible between animation frames. Thus, simulations running on faster computers may take smaller steps and be more accurate. The need for eventual consistency precludes this strategy here. Instead, the desired step length for \ensuremath{\Varid{collaborationOf}} is defined globally and set to one-sixteenth of a second:
\begin{hscode}\SaveRestoreHook
\column{B}{@{}>{\hspre}l<{\hspost}@{}}%
\column{E}{@{}>{\hspre}l<{\hspost}@{}}%
\>[B]{}\Varid{gameRate}\mathbin{::}\Conid{Double}{}\<[E]%
\\[-0.3ex]%
\>[B]{}\Varid{gameRate}\mathrel{=}\mathrm{1}\mathbin{/}\mathrm{16}{}\<[E]%
\ColumnHook
\end{hscode}\resethooks

We can obtain the desired resolution by wrapping the student's step function in one that iterates \ensuremath{\Varid{step}} on time steps larger than the desired rate:

\begin{hscode}\SaveRestoreHook
\column{B}{@{}>{\hspre}l<{\hspost}@{}}%
\column{20}{@{}>{\hspre}c<{\hspost}@{}}%
\column{20E}{@{}l@{}}%
\column{23}{@{}>{\hspre}l<{\hspost}@{}}%
\column{38}{@{}>{\hspre}l<{\hspost}@{}}%
\column{E}{@{}>{\hspre}l<{\hspost}@{}}%
\>[B]{}\Varid{gameStep}\mathbin{::}\Conid{Game}\;\Varid{world}\Rightarrow \Conid{Double}\hsarrow{\rightarrow }{\mathpunct{.}}\Varid{world}\hsarrow{\rightarrow }{\mathpunct{.}}\Varid{world}{}\<[E]%
\\[-0.3ex]%
\>[B]{}\Varid{gameStep}\;\Varid{dt}\;\Varid{world}{}\<[20]%
\>[20]{}\mid {}\<[20E]%
\>[23]{}\Varid{dt}\leq \mathrm{0}{}\<[38]%
\>[38]{}\mathrel{=}\Varid{world}{}\<[E]%
\\[-0.3ex]%
\>[20]{}\mid {}\<[20E]%
\>[23]{}\Varid{dt}\mathbin{>}\Varid{gameRate}{}\<[38]%
\>[38]{}\mathrel{=}\Varid{gameStep}\;(\Varid{dt}\mathbin{-}\Varid{gameRate})\;(\Varid{step}\;\Varid{gameRate}\;\Varid{world}){}\<[E]%
\\[-0.3ex]%
\>[20]{}\mid {}\<[20E]%
\>[23]{}\Varid{otherwise}{}\<[38]%
\>[38]{}\mathrel{=}\Varid{step}\;\Varid{dt}\;\Varid{world}{}\<[E]%
\ColumnHook
\end{hscode}\resethooks
Replacing \ensuremath{\Varid{step}} with \ensuremath{\Varid{gameStep}} in the implementation of \ensuremath{\Varid{currentState}} and \ensuremath{\Varid{applyEvents}} above yields a correct solution.

To see this code in action, we construct the following program: As the time passes, a column grows on the screen, from bottom to top. Initially, it is gray. When a player presses a number key, the column begins to grow in a different color. Additionally, whenever \ensuremath{\Varid{step}} is called, this current height of the column is marked with a black line.

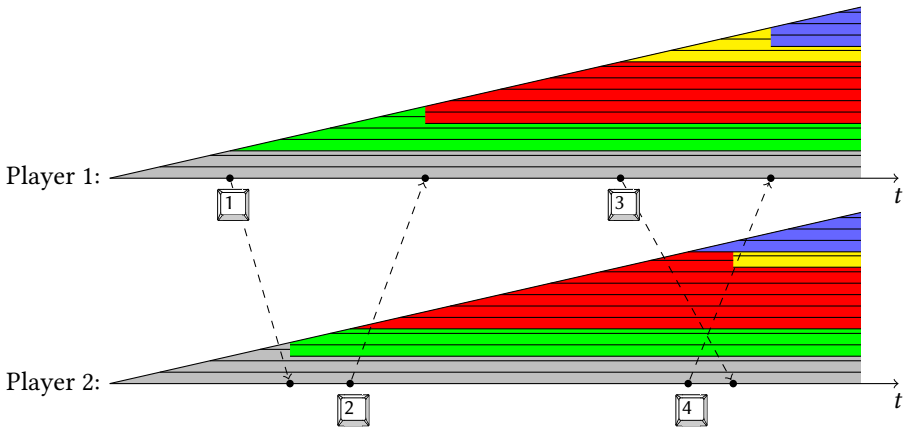
\begin{figure}
\centering
\begin{tikzpicture}[xscale=2,yscale=0.9]
\makeatletter
\newcommand\currentcoordinate{\the\tikz@lastxsaved,\the\tikz@lastysaved}
\makeatother

\def\s{0.5}
\def\E{5}

\def\A{0.8}
\def\Ar{1.2}
\def\B{1.6}
\def\Br{2.1}
\def\C{3.4}
\def\Cr{4.15}
\def\D{3.85}
\def\Dr{4.4}

\fill[color=gray!50] (0,0) -- (\E,0) -- (\E,\E*\s) -- cycle;
\fill[color=green]   (\A,\A*\s) -- (\E,\A*\s) -- (\E,\E*\s) -- cycle;
\fill[color=red]     (\Br,\B*\s) -- (\E,\B*\s) -- (\E,\E*\s) -- (\Br,\Br*\s) -- cycle;
\fill[color=yellow]  (\C,\C*\s) -- (\E,\C*\s) -- (\E,\E*\s) -- cycle;
\fill[color=blue!60] (\Dr,\D*\s) -- (\E,\D*\s) -- (\E,\E*\s) -- (\Dr,\Dr*\s) -- cycle;
\draw[->] (0,0) node [left] {Player 1:} -- (\E,0) -- ++(0.25,0) node [below] {$t$};

\draw (0,0) -- (\E,\E*\s);
\draw (1/3,1/3*\s) -- (\E,1/3*\s);
\draw (2/3,2/3*\s) -- (\E,2/3*\s);

\draw ($(\A,\A*\s)+(0/3,0/3*\s)$) -- ($(\E,\A*\s)+(0,0/3*\s)$);
\draw ($(\A,\A*\s)+(1/3,1/3*\s)$) -- ($(\E,\A*\s)+(0,1/3*\s)$);
\draw ($(\A,\A*\s)+(2/3,2/3*\s)$) -- ($(\E,\A*\s)+(0,2/3*\s)$);
\draw ($(\A,\A*\s)+(3/3,3/3*\s)$) -- ($(\Br,\A*\s)+(0,3/3*\s)$);

\draw ($(\Br,\B*\s)+(0   ,0/3*\s)$) -- ($(\E,\B*\s)+(0,0/3*\s)$);
\draw ($(\Br,\B*\s)+(0   ,1/3*\s)$) -- ($(\E,\B*\s)+(0,1/3*\s)$);
\draw ($(\B, \B*\s)+(2/3 ,2/3*\s)$) -- ($(\E,\B*\s)+(0,2/3*\s)$);
\draw ($(\B, \B*\s)+(3/3 ,3/3*\s)$) -- ($(\E,\B*\s)+(0,3/3*\s)$);
\draw ($(\B, \B*\s)+(4/3 ,4/3*\s)$) -- ($(\E,\B*\s)+(0,4/3*\s)$);
\draw ($(\B, \B*\s)+(5/3 ,5/3*\s)$) -- ($(\E,\B*\s)+(0,5/3*\s)$);

\draw ($(\C,\C*\s)+(0/3,0/3*\s)$) -- ($(\E, \C*\s)+(0,0/3*\s)$);
\draw ($(\C,\C*\s)+(1/3,1/3*\s)$) -- ($(\E, \C*\s)+(0,1/3*\s)$);
\draw ($(\C,\C*\s)+(2/3,2/3*\s)$) -- ($(\Dr,\C*\s)+(0,2/3*\s)$);

\draw ($(\Dr,\D*\s)+(0   ,0/3*\s)$) -- ($(\E,\D*\s)+(0,0/3*\s)$);
\draw ($(\Dr,\D*\s)+(0   ,1/3*\s)$) -- ($(\E,\D*\s)+(0,1/3*\s)$);
\draw ($(\D, \D*\s)+(2/3 ,2/3*\s)$) -- ($(\E,\D*\s)+(0,2/3*\s)$);
\draw ($(\D, \D*\s)+(3/3 ,3/3*\s)$) -- ($(\E,\D*\s)+(0,3/3*\s)$);

\node[inner sep=1pt, fill, circle] (A)  at (\A, 0) {};
\node[inner sep=1pt, fill, circle] (Br) at (\Br,0) {};
\node[inner sep=1pt, fill, circle] (C)  at (\C ,0) {};
\node[inner sep=1pt, fill, circle] (Dr) at (\Dr,0) {};

\begin{scope}[shift={(0,-6*\s)}]
\fill[color=gray!50] (0,0) -- (\E,0) -- (\E,\E*\s) -- cycle;
\fill[color=green]   (\Ar,\A*\s) -- (\E,\A*\s) -- (\E,\E*\s) -- (\Ar,\Ar*\s) -- cycle;
\fill[color=red]     (\B,\B*\s) -- (\E,\B*\s) -- (\E,\E*\s) -- cycle;
\fill[color=yellow]  (\Cr,\C*\s) -- (\E,\C*\s) -- (\E,\E*\s) -- (\Cr,\Cr*\s) -- cycle;
\fill[color=blue!60] (\D,\D*\s) -- (\E,\D*\s) -- (\E,\E*\s) -- cycle;
\draw[->] (0,0) node [left] {Player 2:} -- (\E,0) -- ++(0.25,0) node [below] {$t$};

\draw (0,0) -- (\E,\E*\s);
\draw (1/3,1/3*\s) -- (\E,1/3*\s);
\draw (2/3,2/3*\s) -- (\E,2/3*\s);
\draw (3/3,3/3*\s) -- (\Ar,3/3*\s);

\draw ($(\Ar,\A*\s)+(0  ,0/3*\s)$) -- ($(\E,\A*\s)+(0,0/3*\s)$);
\draw ($(\Ar,\A*\s)+(0  ,1/3*\s)$) -- ($(\E,\A*\s)+(0,1/3*\s)$);
\draw ($(\A,\A*\s)+(2/3,2/3*\s)$) -- ($(\E,\A*\s)+(0,2/3*\s)$);

\draw ($(\B,\B*\s)+(0/3 ,0/3*\s)$) -- ($(\E,\B*\s)+(0,0/3*\s)$);
\draw ($(\B,\B*\s)+(1/3 ,1/3*\s)$) -- ($(\E,\B*\s)+(0,1/3*\s)$);
\draw ($(\B,\B*\s)+(2/3 ,2/3*\s)$) -- ($(\E,\B*\s)+(0,2/3*\s)$);
\draw ($(\B,\B*\s)+(3/3 ,3/3*\s)$) -- ($(\E,\B*\s)+(0,3/3*\s)$);
\draw ($(\B,\B*\s)+(4/3 ,4/3*\s)$) -- ($(\E,\B*\s)+(0,4/3*\s)$);
\draw ($(\B,\B*\s)+(5/3 ,5/3*\s)$) -- ($(\E,\B*\s)+(0,5/3*\s)$);
\draw ($(\B,\B*\s)+(6/3 ,6/3*\s)$) -- ($(\Cr,\B*\s)+(0,6/3*\s)$);

\draw ($(\Cr,\C*\s)+(0   ,0/3*\s)$) -- ($(\E,\C*\s)+(0,0/3*\s)$);
\draw ($(\Cr,\C*\s)+(0   ,1/3*\s)$) -- ($(\E,\C*\s)+(0,1/3*\s)$);

\draw ($(\D, \D*\s)+(0/3 ,0/3*\s)$) -- ($(\E,\D*\s)+(0,0/3*\s)$);
\draw ($(\D, \D*\s)+(1/3 ,1/3*\s)$) -- ($(\E,\D*\s)+(0,1/3*\s)$);
\draw ($(\D, \D*\s)+(2/3 ,2/3*\s)$) -- ($(\E,\D*\s)+(0,2/3*\s)$);
\draw ($(\D, \D*\s)+(3/3 ,3/3*\s)$) -- ($(\E,\D*\s)+(0,3/3*\s)$);

\node[inner sep=1pt, fill, circle] (Ar) at (\Ar,0) {};
\node[inner sep=1pt, fill, circle] (B)  at (\B, 0) {};
\node[inner sep=1pt, fill, circle] (Cr) at (\Cr,0) {};
\node[inner sep=1pt, fill, circle] (D)  at (\D, 0) {};
\end{scope}

\draw[dashed, ->] (A) -- (Ar);
\draw[dashed, ->] (B) -- (Br);
\draw[dashed, ->] (C) -- (Cr);
\draw[dashed, ->] (D) -- (Dr);

\node[below,fill=white,inner sep=0pt,outer sep=4pt] at (A) {\keystroke{1}};
\node[below,fill=white,inner sep=0pt,outer sep=4pt] at (B) {\keystroke{2}};
\node[below,fill=white,inner sep=0pt,outer sep=4pt] at (C) {\keystroke{3}};
\node[below,fill=white,inner sep=0pt,outer sep=4pt] at (D) {\keystroke{4}};

\end{tikzpicture}
\caption{The evolution of a simple multi-player program with latency in the messages}
\label{fig:evolution}
\end{figure}

Because the program output is one-dimensional, we can use the horizontal dimension to show in \cref{fig:evolution} how the players' displays evolves over time. The dashed arrows indicate the transfer of each packet to the other player, which is not instant. When a message from the other player arrives, the state is updated to reflect this change. Because this game essentially records its history, these delayed updates result in a ``flicker" as the client updates the state. In many cases the effect will be less noticeable than it is here. We can see that the algorithm achieved eventual consistency, as the right edge of the drawing looks identical for both clients.

\subsection{Limiting time travel}
\label{sec:committing}

In the course of a game, quite a large number of events occur. As time goes by, the cost of calculating the current state from scratch grows without bound, and will eventually become too large to be completed between each frame, and animations will stop being smooth. Clearly, some of that computation is quite pointless to repeat.

Our message transport guarantees that messages from each client are delivered in order, so that when we receive a message, we know that we have seen all messages from the sender up to that timestamp. If we call this the client's commit time, then we know that no new events will be received before the earliest commit time of any client, which we call the commit horizon. We can now precompute the game state up to the commit horizon, forget all older state and events, and use this as the basis for future state recalculations.

In the following we will explain the data structure and associated operations that CodeWorld  uses to keep track of the committed state, the pending events and each player's commit time. The main data type is
\begin{hscode}\SaveRestoreHook
\column{B}{@{}>{\hspre}l<{\hspost}@{}}%
\column{23}{@{}>{\hspre}c<{\hspost}@{}}%
\column{23E}{@{}l@{}}%
\column{26}{@{}>{\hspre}l<{\hspost}@{}}%
\column{37}{@{}>{\hspre}c<{\hspost}@{}}%
\column{37E}{@{}l@{}}%
\column{41}{@{}>{\hspre}l<{\hspost}@{}}%
\column{E}{@{}>{\hspre}l<{\hspost}@{}}%
\>[B]{}\textbf{\lmss data}\;\Conid{Log}\;\Varid{world}\mathrel{=}\Conid{Log}\;{}\<[23]%
\>[23]{}\{\mskip1.5mu {}\<[23E]%
\>[26]{}\Varid{committed}{}\<[37]%
\>[37]{}\mathbin{::}{}\<[37E]%
\>[41]{}(\Conid{Timestamp},\Varid{world}),{}\<[E]%
\\[-0.3ex]%
\>[26]{}\Varid{events}{}\<[37]%
\>[37]{}\mathbin{::}{}\<[37E]%
\>[41]{}[\mskip1.5mu \Conid{Message}\mskip1.5mu],{}\<[E]%
\\[-0.3ex]%
\>[26]{}\Varid{latest}{}\<[37]%
\>[37]{}\mathbin{::}{}\<[37E]%
\>[41]{}[\mskip1.5mu (\Conid{Player},\Conid{Timestamp})\mskip1.5mu]\mskip1.5mu\}{}\<[E]%
\ColumnHook
\end{hscode}\resethooks

Initially, there are no events, and everything is at timestamp zero:
\begin{hscode}\SaveRestoreHook
\column{B}{@{}>{\hspre}l<{\hspost}@{}}%
\column{E}{@{}>{\hspre}l<{\hspost}@{}}%
\>[B]{}\Varid{initLog}\mathbin{::}\Conid{Game}\;\Varid{world}\Rightarrow [\mskip1.5mu \Conid{Player}\mskip1.5mu]\hsarrow{\rightarrow }{\mathpunct{.}}\Conid{Log}\;\Varid{world}{}\<[E]%
\\[-0.3ex]%
\>[B]{}\Varid{initLog}\;\Varid{ps}\mathrel{=}\Conid{Log}\;(\mathrm{0},\Varid{start})\;[\mskip1.5mu \mskip1.5mu]\;[\mskip1.5mu (\Varid{p},\mathrm{0})\mid \Varid{p}\leftarrow \Varid{ps}\mskip1.5mu]{}\<[E]%
\ColumnHook
\end{hscode}\resethooks
When an event comes in, the message is added to \ensuremath{\Varid{events}} via the public \ensuremath{\Varid{addEvent}} function.
\begin{hscode}\SaveRestoreHook
\column{B}{@{}>{\hspre}l<{\hspost}@{}}%
\column{3}{@{}>{\hspre}l<{\hspost}@{}}%
\column{10}{@{}>{\hspre}l<{\hspost}@{}}%
\column{E}{@{}>{\hspre}l<{\hspost}@{}}%
\>[B]{}\Varid{addEvent}\mathbin{::}\Conid{Game}\;\Varid{world}\Rightarrow \Conid{Message}\hsarrow{\rightarrow }{\mathpunct{.}}\Conid{Log}\;\Varid{world}\hsarrow{\rightarrow }{\mathpunct{.}}\Conid{Log}\;\Varid{world}{}\<[E]%
\\[-0.3ex]%
\>[B]{}\Varid{addEvent}\;(\Varid{t},\Varid{p},\Varid{e})\;\Varid{log}\mathrel{=}\Varid{recordActivity}\;\Varid{t}\;\Varid{p}\;(\Varid{log}\;\{\mskip1.5mu \Varid{events}\mathrel{=}\Varid{events'}\mskip1.5mu\}){}\<[E]%
\\[-0.3ex]%
\>[B]{}\hsindent{3}{}\<[3]%
\>[3]{}\textbf{\lmss where}\;{}\<[10]%
\>[10]{}\Varid{events'}\mathrel{=}\Varid{sortMessages}\;(\Varid{events}\;\Varid{log}\mathbin{+\!+}[\mskip1.5mu (\Varid{t},\Varid{p},\Varid{e})\mskip1.5mu]){}\<[E]%
\ColumnHook
\end{hscode}\resethooks
Then, the client's commit time in \ensuremath{\Varid{latest}} is updated.
\begin{hscode}\SaveRestoreHook
\column{B}{@{}>{\hspre}l<{\hspost}@{}}%
\column{3}{@{}>{\hspre}l<{\hspost}@{}}%
\column{10}{@{}>{\hspre}l<{\hspost}@{}}%
\column{26}{@{}>{\hspre}c<{\hspost}@{}}%
\column{26E}{@{}l@{}}%
\column{29}{@{}>{\hspre}l<{\hspost}@{}}%
\column{44}{@{}>{\hspre}l<{\hspost}@{}}%
\column{E}{@{}>{\hspre}l<{\hspost}@{}}%
\>[B]{}\Varid{recordActivity}\mathbin{::}\Conid{Game}\;\Varid{world}\Rightarrow \Conid{Timestamp}\hsarrow{\rightarrow }{\mathpunct{.}}\Conid{Player}\hsarrow{\rightarrow }{\mathpunct{.}}\Conid{Log}\;\Varid{world}\hsarrow{\rightarrow }{\mathpunct{.}}\Conid{Log}\;\Varid{world}{}\<[E]%
\\[-0.3ex]%
\>[B]{}\Varid{recordActivity}\;\Varid{t}\;\Varid{p}\;\Varid{log}{}\<[26]%
\>[26]{}\mid {}\<[26E]%
\>[29]{}\Varid{t}\mathbin{<}\Varid{t\char95 old}{}\<[44]%
\>[44]{}\mathrel{=}\Varid{error}\;\text{\itshape\texttt{\char34}\!Messages~out~of~order\texttt{\char34}}{}\<[E]%
\\[-0.3ex]%
\>[26]{}\mid {}\<[26E]%
\>[29]{}\Varid{otherwise}{}\<[44]%
\>[44]{}\mathrel{=}\Varid{advanceCommitted}\;(\Varid{log}\;\{\mskip1.5mu \Varid{latest}\mathrel{=}\Varid{latest'}\mskip1.5mu\}){}\<[E]%
\\[-0.3ex]%
\>[B]{}\hsindent{3}{}\<[3]%
\>[3]{}\textbf{\lmss where}\;{}\<[10]%
\>[10]{}\Varid{latest'}\mathrel{=}(\Varid{p},\Varid{t})\mathbin{:}\Varid{delete}\;(\Varid{p},\Varid{t\char95 old})\;(\Varid{latest}\;\Varid{log}){}\<[E]%
\\[-0.3ex]%
\>[10]{}\Conid{Just}\;\Varid{t\char95 old}\mathrel{=}\Varid{lookup}\;\Varid{p}\;(\Varid{latest}\;\Varid{log}){}\<[E]%
\ColumnHook
\end{hscode}\resethooks
This might have moved the commit horizon, and if some of the messages from the list \ensuremath{\Varid{events}} are from before the commit horizon, we can integrate them into the \ensuremath{\Varid{committed}} state.
\begin{hscode}\SaveRestoreHook
\column{B}{@{}>{\hspre}l<{\hspost}@{}}%
\column{3}{@{}>{\hspre}l<{\hspost}@{}}%
\column{10}{@{}>{\hspre}l<{\hspost}@{}}%
\column{29}{@{}>{\hspre}c<{\hspost}@{}}%
\column{29E}{@{}l@{}}%
\column{32}{@{}>{\hspre}l<{\hspost}@{}}%
\column{43}{@{}>{\hspre}l<{\hspost}@{}}%
\column{E}{@{}>{\hspre}l<{\hspost}@{}}%
\>[B]{}\Varid{advanceCommitted}\mathbin{::}\Conid{Game}\;\Varid{world}\Rightarrow \Conid{Log}\;\Varid{world}\hsarrow{\rightarrow }{\mathpunct{.}}\Conid{Log}\;\Varid{world}{}\<[E]%
\\[-0.3ex]%
\>[B]{}\Varid{advanceCommitted}\;\Varid{log}\mathrel{=}\Varid{log}\;{}\<[29]%
\>[29]{}\{\mskip1.5mu {}\<[29E]%
\>[32]{}\Varid{events}{}\<[43]%
\>[43]{}\mathrel{=}\Varid{to\char95 keep},{}\<[E]%
\\[-0.3ex]%
\>[32]{}\Varid{committed}{}\<[43]%
\>[43]{}\mathrel{=}\Varid{applyEvents}\;\Varid{to\char95 commit}\;(\Varid{committed}\;\Varid{log})\mskip1.5mu\}{}\<[E]%
\\[-0.3ex]%
\>[B]{}\hsindent{3}{}\<[3]%
\>[3]{}\textbf{\lmss where}\;{}\<[10]%
\>[10]{}(\Varid{to\char95 commit},\Varid{to\char95 keep})\mathrel{=}\Varid{span}\;(\mathbf{\lmss\uplambda}\hslambda (\Varid{t},\anonymous ,\anonymous )\hsarrow{\rightarrow }{\mathpunct{.}}\Varid{t}\mathbin{<}\Varid{commitHorizon}\;\Varid{log})\;(\Varid{events}\;\Varid{log}){}\<[E]%
\\[\blanklineskip]%
\>[B]{}\Varid{commitHorizon}\mathbin{::}\Conid{Log}\;\Varid{world}\hsarrow{\rightarrow }{\mathpunct{.}}\Conid{Timestamp}{}\<[E]%
\\[-0.3ex]%
\>[B]{}\Varid{commitHorizon}\;\Varid{log}\mathrel{=}\Varid{minimum}\;[\mskip1.5mu \Varid{t}\mid (\Varid{p},\Varid{t})\leftarrow \Varid{latest}\;\Varid{log}\mskip1.5mu]{}\<[E]%
\ColumnHook
\end{hscode}\resethooks
The final public function is used to query the current state of the game. Starting from the committed state, it applies the pending events.
\begin{hscode}\SaveRestoreHook
\column{B}{@{}>{\hspre}l<{\hspost}@{}}%
\column{3}{@{}>{\hspre}l<{\hspost}@{}}%
\column{10}{@{}>{\hspre}l<{\hspost}@{}}%
\column{E}{@{}>{\hspre}l<{\hspost}@{}}%
\>[B]{}\Varid{currentState}\mathbin{::}\Conid{Game}\;\Varid{world}\Rightarrow \Conid{Timestamp}\hsarrow{\rightarrow }{\mathpunct{.}}\Conid{Log}\;\Varid{world}\hsarrow{\rightarrow }{\mathpunct{.}}\Varid{world}{}\<[E]%
\\[-0.3ex]%
\>[B]{}\Varid{currentState}\;\Varid{now}\;\Varid{log}\mid \Varid{now}\mathbin{<}\Varid{commitHorizon}\;\Varid{log}\mathrel{=}\Varid{error}\;\text{\itshape\texttt{\char34}\!Cannot~look~into~the~past\texttt{\char34}}{}\<[E]%
\\[-0.3ex]%
\>[B]{}\Varid{currentState}\;\Varid{now}\;\Varid{log}\mathrel{=}\Varid{gameStep}\;(\Varid{now}\mathbin{-}\Varid{t})\;\Varid{world}{}\<[E]%
\\[-0.3ex]%
\>[B]{}\hsindent{3}{}\<[3]%
\>[3]{}\textbf{\lmss where}\;{}\<[10]%
\>[10]{}\Varid{past\char95 events}\mathrel{=}\Varid{takeWhile}\;(\mathbf{\lmss\uplambda}\hslambda (\Varid{t},\anonymous ,\anonymous )\hsarrow{\rightarrow }{\mathpunct{.}}\Varid{t}\leq \Varid{now})\;(\Varid{events}\;\Varid{log}){}\<[E]%
\\[-0.3ex]%
\>[10]{}(\Varid{t},\Varid{world})\mathrel{=}\Varid{applyEvents}\;\Varid{past\char95 events}\;(\Varid{committed}\;\Varid{log}){}\<[E]%
\ColumnHook
\end{hscode}\resethooks

This algorithm\iflong, printed in \cref{fig:committing} in its entirety,\fi{} relies on these assumptions:
\begin{compactenum}
\item The list of players provided to \ensuremath{\Varid{initLog}} is correct.
\item For each player, events are added in order, with monotonically increasing timestamps.
\item The state is never queried at a time that lies before \ensuremath{\Varid{commitHorizon}}.
\end{compactenum}

The first assumption is ensured by the CodeWorld framework. The second is ensured by using a monotonic time source to create the timestamps, and by using an order-preserving communication channel. The third follows from the fact that every client's own timestamps are always in that player’s past, and therefore the argument to \ensuremath{\Varid{currentState}} is later than the commit horizon.

If one of the players were to stop interacting with the program, that client would not send any messages. In this case, no events can be committed and the list of events to be processed by \ensuremath{\Varid{currentState}} would again grow without bound. To avoid this, each client sends empty messages (``pings") whenever the user has not produced input for a certain amount of time. When such a ping is received, the \ensuremath{\Varid{addPing}} function advances the \ensuremath{\Varid{latest}} field without adding a new event:
\begin{hscode}\SaveRestoreHook
\column{B}{@{}>{\hspre}l<{\hspost}@{}}%
\column{E}{@{}>{\hspre}l<{\hspost}@{}}%
\>[B]{}\Varid{addPing}\mathbin{::}\Conid{Game}\;\Varid{world}\Rightarrow (\Conid{Timestamp},\Conid{Player})\hsarrow{\rightarrow }{\mathpunct{.}}\Conid{Log}\;\Varid{world}\hsarrow{\rightarrow }{\mathpunct{.}}\Conid{Log}\;\Varid{world}{}\<[E]%
\\[-0.3ex]%
\>[B]{}\Varid{addPing}\;(\Varid{t},\Varid{p})\;\Varid{log}\mathrel{=}\Varid{recordActivity}\;\Varid{t}\;\Varid{p}\;\Varid{log}{}\<[E]%
\ColumnHook
\end{hscode}\resethooks
\iflong

This way, the number of events in the \ensuremath{\Varid{events}} field is bound by
\[
\text{max input rate} \times (\text{max network delay} + \text{max time between events or pings}) \times (\text{number of players}-1)
\]
which is independent of how long the game has been running.\jb{As long as no players disconnect. Not sure how best to phrase that. Maybe not important. But one reviewer noticed. Worth a sentence?}
\else
Assuming a bounded input event rate and network delay, this bounds the size of the \ensuremath{\Varid{events}} field.
\fi

More tweaks are possible. In the CodeWorld implementation, we also cache the current state, so that querying the current state again, when no new events were received, is much cheaper. When an input event from another player comes in, we discard this cached value and recalculate it based on the committed state and the stored events.

\iflong
\begin{figure}
\begin{hscode}\SaveRestoreHook
\column{B}{@{}>{\hspre}l<{\hspost}@{}}%
\column{3}{@{}>{\hspre}l<{\hspost}@{}}%
\column{10}{@{}>{\hspre}l<{\hspost}@{}}%
\column{20}{@{}>{\hspre}c<{\hspost}@{}}%
\column{20E}{@{}l@{}}%
\column{23}{@{}>{\hspre}l<{\hspost}@{}}%
\column{24}{@{}>{\hspre}c<{\hspost}@{}}%
\column{24E}{@{}l@{}}%
\column{26}{@{}>{\hspre}c<{\hspost}@{}}%
\column{26E}{@{}l@{}}%
\column{27}{@{}>{\hspre}l<{\hspost}@{}}%
\column{29}{@{}>{\hspre}c<{\hspost}@{}}%
\column{29E}{@{}l@{}}%
\column{30}{@{}>{\hspre}l<{\hspost}@{}}%
\column{32}{@{}>{\hspre}l<{\hspost}@{}}%
\column{38}{@{}>{\hspre}c<{\hspost}@{}}%
\column{38E}{@{}l@{}}%
\column{39}{@{}>{\hspre}l<{\hspost}@{}}%
\column{42}{@{}>{\hspre}l<{\hspost}@{}}%
\column{43}{@{}>{\hspre}l<{\hspost}@{}}%
\column{44}{@{}>{\hspre}l<{\hspost}@{}}%
\column{E}{@{}>{\hspre}l<{\hspost}@{}}%
\>[B]{}\textbf{\lmss type}\;\Conid{Timestamp}{}\<[20]%
\>[20]{}\mathrel{=}{}\<[20E]%
\>[23]{}\Conid{Double}{}\<[E]%
\\[-0.3ex]%
\>[B]{}\textbf{\lmss type}\;\Conid{Player}{}\<[20]%
\>[20]{}\mathrel{=}{}\<[20E]%
\>[23]{}\Conid{Int}{}\<[E]%
\\[-0.3ex]%
\>[B]{}\textbf{\lmss type}\;\Conid{Message}{}\<[20]%
\>[20]{}\mathrel{=}{}\<[20E]%
\>[23]{}(\Conid{Timestamp},\Conid{Player},\Conid{Event}){}\<[E]%
\\[-0.3ex]%
\>[B]{}\textbf{\lmss data}\;\Conid{Log}\;\Varid{world}\mathrel{=}\Conid{Log}\;{}\<[24]%
\>[24]{}\{\mskip1.5mu {}\<[24E]%
\>[27]{}\Varid{committed}{}\<[38]%
\>[38]{}\mathbin{::}{}\<[38E]%
\>[42]{}(\Conid{Timestamp},\Varid{world}),{}\<[E]%
\\[-0.3ex]%
\>[27]{}\Varid{events}{}\<[38]%
\>[38]{}\mathbin{::}{}\<[38E]%
\>[42]{}[\mskip1.5mu \Conid{Message}\mskip1.5mu],{}\<[E]%
\\[-0.3ex]%
\>[27]{}\Varid{latest}{}\<[38]%
\>[38]{}\mathbin{::}{}\<[38E]%
\>[42]{}[\mskip1.5mu (\Conid{Player},\Conid{Timestamp})\mskip1.5mu]\mskip1.5mu\}{}\<[E]%
\\[\blanklineskip]%
\>[B]{}\mbox{\onelinecomment  Public interface}{}\<[E]%
\\[-0.3ex]%
\>[B]{}\Varid{initLog}\mathbin{::}\Conid{Game}\;\Varid{world}\Rightarrow [\mskip1.5mu \Conid{Player}\mskip1.5mu]\hsarrow{\rightarrow }{\mathpunct{.}}\Conid{Log}\;\Varid{world}{}\<[E]%
\\[-0.3ex]%
\>[B]{}\Varid{initLog}\;\Varid{ps}\mathrel{=}\Conid{Log}\;(\mathrm{0},\Varid{start})\;[\mskip1.5mu \mskip1.5mu]\;([\mskip1.5mu (\Varid{p},\mathrm{0})\mid \Varid{p}\leftarrow \Varid{ps}\mskip1.5mu]){}\<[E]%
\\[\blanklineskip]%
\>[B]{}\Varid{addEvent}\mathbin{::}\Conid{Game}\;\Varid{world}\Rightarrow \Conid{Message}\hsarrow{\rightarrow }{\mathpunct{.}}\Conid{Log}\;\Varid{world}\hsarrow{\rightarrow }{\mathpunct{.}}\Conid{Log}\;\Varid{world}{}\<[E]%
\\[-0.3ex]%
\>[B]{}\Varid{addEvent}\;(\Varid{t},\Varid{p},\Varid{e})\;\Varid{log}\mathrel{=}\Varid{recordActivity}\;\Varid{t}\;\Varid{p}\;(\Varid{log}\;\{\mskip1.5mu \Varid{events}\mathrel{=}\Varid{events'}\mskip1.5mu\}){}\<[E]%
\\[-0.3ex]%
\>[B]{}\hsindent{3}{}\<[3]%
\>[3]{}\textbf{\lmss where}\;{}\<[10]%
\>[10]{}\Varid{events'}\mathrel{=}\Varid{sortMessages}\;(\Varid{events}\;\Varid{log}\mathbin{+\!+}[\mskip1.5mu (\Varid{t},\Varid{p},\Varid{e})\mskip1.5mu]){}\<[E]%
\\[\blanklineskip]%
\>[B]{}\Varid{addPing}\mathbin{::}\Conid{Game}\;\Varid{world}\Rightarrow (\Conid{Timestamp},\Conid{Player})\hsarrow{\rightarrow }{\mathpunct{.}}\Conid{Log}\;\Varid{world}\hsarrow{\rightarrow }{\mathpunct{.}}\Conid{Log}\;\Varid{world}{}\<[E]%
\\[-0.3ex]%
\>[B]{}\Varid{addPing}\;(\Varid{t},\Varid{p})\;\Varid{log}\mathrel{=}\Varid{recordActivity}\;\Varid{t}\;\Varid{p}\mathbin{\$}\Varid{log}{}\<[E]%
\\[\blanklineskip]%
\>[B]{}\Varid{currentState}\mathbin{::}\Conid{Game}\;\Varid{world}\Rightarrow \Conid{Timestamp}\hsarrow{\rightarrow }{\mathpunct{.}}\Conid{Log}\;\Varid{world}\hsarrow{\rightarrow }{\mathpunct{.}}\Varid{world}{}\<[E]%
\\[-0.3ex]%
\>[B]{}\Varid{currentState}\;\Varid{now}\;\Varid{log}\mid \Varid{now}\mathbin{<}\Varid{commitHorizon}\;\Varid{log}\mathrel{=}\Varid{error}\;\text{\itshape\texttt{\char34}\!Cannot~look~into~the~past\texttt{\char34}}{}\<[E]%
\\[-0.3ex]%
\>[B]{}\Varid{currentState}\;\Varid{now}\;\Varid{log}\mathrel{=}\Varid{gameStep}\;(\Varid{now}\mathbin{-}\Varid{t})\;\Varid{world}{}\<[E]%
\\[-0.3ex]%
\>[B]{}\hsindent{3}{}\<[3]%
\>[3]{}\textbf{\lmss where}\;{}\<[10]%
\>[10]{}\Varid{past\char95 events}\mathrel{=}\Varid{takeWhile}\;(\mathbf{\lmss\uplambda}\hslambda (\Varid{t},\anonymous ,\anonymous )\hsarrow{\rightarrow }{\mathpunct{.}}\Varid{t}\leq \Varid{now})\;(\Varid{events}\;\Varid{log}){}\<[E]%
\\[-0.3ex]%
\>[10]{}(\Varid{t},\Varid{world})\mathrel{=}\Varid{applyEvents}\;\Varid{past\char95 events}\;(\Varid{committed}\;\Varid{log}){}\<[E]%
\\[\blanklineskip]%
\>[B]{}\mbox{\onelinecomment  Internal functions}{}\<[E]%
\\[-0.3ex]%
\>[B]{}\Varid{gameRate}\mathbin{::}\Conid{Double}{}\<[E]%
\\[-0.3ex]%
\>[B]{}\Varid{gameRate}\mathrel{=}\mathrm{1}\mathbin{/}\mathrm{16}{}\<[E]%
\\[\blanklineskip]%
\>[B]{}\Varid{sortMessages}\mathbin{::}[\mskip1.5mu \Conid{Message}\mskip1.5mu]\hsarrow{\rightarrow }{\mathpunct{.}}[\mskip1.5mu \Conid{Message}\mskip1.5mu]{}\<[E]%
\\[-0.3ex]%
\>[B]{}\Varid{sortMessages}\mathrel{=}\Varid{sortOn}\;(\mathbf{\lmss\uplambda}\hslambda (\Varid{t},\Varid{p},\anonymous )\hsarrow{\rightarrow }{\mathpunct{.}}(\Varid{t},\Varid{p})){}\<[E]%
\\[\blanklineskip]%
\>[B]{}\Varid{recordActivity}\mathbin{::}\Conid{Game}\;\Varid{world}\Rightarrow \Conid{Timestamp}\hsarrow{\rightarrow }{\mathpunct{.}}\Conid{Player}\hsarrow{\rightarrow }{\mathpunct{.}}\Conid{Log}\;\Varid{world}\hsarrow{\rightarrow }{\mathpunct{.}}\Conid{Log}\;\Varid{world}{}\<[E]%
\\[-0.3ex]%
\>[B]{}\Varid{recordActivity}\;\Varid{t}\;\Varid{p}\;\Varid{log}{}\<[26]%
\>[26]{}\mid {}\<[26E]%
\>[30]{}\Varid{t}\mathbin{<}\Varid{t\char95 old}{}\<[44]%
\>[44]{}\mathrel{=}\Varid{error}\;\text{\itshape\texttt{\char34}\!Messages~out~of~order\texttt{\char34}}{}\<[E]%
\\[-0.3ex]%
\>[26]{}\mid {}\<[26E]%
\>[30]{}\Varid{otherwise}{}\<[44]%
\>[44]{}\mathrel{=}\Varid{advanceCommitted}\;(\Varid{log}\;\{\mskip1.5mu \Varid{latest}\mathrel{=}\Varid{latest'}\mskip1.5mu\}){}\<[E]%
\\[-0.3ex]%
\>[B]{}\hsindent{3}{}\<[3]%
\>[3]{}\textbf{\lmss where}\;{}\<[10]%
\>[10]{}\Varid{latest'}\mathrel{=}(\Varid{p},\Varid{t})\mathbin{:}\Varid{delete}\;(\Varid{p},\Varid{t\char95 old})\;(\Varid{latest}\;\Varid{log}){}\<[E]%
\\[-0.3ex]%
\>[10]{}\Conid{Just}\;\Varid{t\char95 old}\mathrel{=}\Varid{lookup}\;\Varid{p}\;(\Varid{latest}\;\Varid{log}){}\<[E]%
\\[\blanklineskip]%
\>[B]{}\Varid{advanceCommitted}\mathbin{::}\Conid{Game}\;\Varid{world}\Rightarrow \Conid{Log}\;\Varid{world}\hsarrow{\rightarrow }{\mathpunct{.}}\Conid{Log}\;\Varid{world}{}\<[E]%
\\[-0.3ex]%
\>[B]{}\Varid{advanceCommitted}\;\Varid{log}\mathrel{=}\Varid{log}\;{}\<[29]%
\>[29]{}\{\mskip1.5mu {}\<[29E]%
\>[32]{}\Varid{committed}{}\<[43]%
\>[43]{}\mathrel{=}\Varid{applyEvents}\;\Varid{to\char95 commit}\;(\Varid{committed}\;\Varid{log}){}\<[E]%
\\[-0.3ex]%
\>[29]{},{}\<[29E]%
\>[32]{}\Varid{events}{}\<[43]%
\>[43]{}\mathrel{=}\Varid{to\char95 keep}\mskip1.5mu\}{}\<[E]%
\\[-0.3ex]%
\>[B]{}\hsindent{3}{}\<[3]%
\>[3]{}\textbf{\lmss where}\;{}\<[10]%
\>[10]{}(\Varid{to\char95 commit},\Varid{to\char95 keep})\mathrel{=}\Varid{span}\;(\mathbf{\lmss\uplambda}\hslambda (\Varid{t},\anonymous ,\anonymous )\hsarrow{\rightarrow }{\mathpunct{.}}\Varid{t}\mathbin{<}\Varid{commitHorizon}\;\Varid{log})\;(\Varid{events}\;\Varid{log}){}\<[E]%
\\[\blanklineskip]%
\>[B]{}\Varid{commitHorizon}\mathbin{::}\Conid{Log}\;\Varid{world}\hsarrow{\rightarrow }{\mathpunct{.}}\Conid{Timestamp}{}\<[E]%
\\[-0.3ex]%
\>[B]{}\Varid{commitHorizon}\;\Varid{log}\mathrel{=}\Varid{minimum}\;[\mskip1.5mu \Varid{t}\mid (\Varid{p},\Varid{t})\leftarrow \Varid{latest}\;\Varid{log}\mskip1.5mu]{}\<[E]%
\\[\blanklineskip]%
\>[B]{}\Varid{applyEvents}\mathbin{::}\Conid{Game}\;\Varid{world}\Rightarrow [\mskip1.5mu \Conid{Message}\mskip1.5mu]\hsarrow{\rightarrow }{\mathpunct{.}}(\Conid{Timestamp},\Varid{world})\hsarrow{\rightarrow }{\mathpunct{.}}(\Conid{Timestamp},\Varid{world}){}\<[E]%
\\[-0.3ex]%
\>[B]{}\Varid{applyEvents}\;\Varid{messages}\;\Varid{ts}\mathrel{=}\Varid{foldl}\;\Varid{apply}\;\Varid{ts}\;\Varid{messages}{}\<[E]%
\\[-0.3ex]%
\>[B]{}\hsindent{3}{}\<[3]%
\>[3]{}\textbf{\lmss where}\;{}\<[10]%
\>[10]{}\Varid{apply}\;(\Varid{t0},\Varid{world})\;(\Varid{t1},\Varid{p},\Varid{e})\mathrel{=}(\Varid{t1},\Varid{handle}\;\Varid{p}\;\Varid{e}\;(\Varid{gameStep}\;(\Varid{t1}\mathbin{-}\Varid{t0})\;\Varid{world})){}\<[E]%
\\[\blanklineskip]%
\>[B]{}\Varid{gameStep}\mathbin{::}\Conid{Game}\;\Varid{world}\Rightarrow \Conid{Double}\hsarrow{\rightarrow }{\mathpunct{.}}\Varid{world}\hsarrow{\rightarrow }{\mathpunct{.}}\Varid{world}{}\<[E]%
\\[-0.3ex]%
\>[B]{}\Varid{gameStep}\;\Varid{dt}\;\Varid{world}{}\<[20]%
\>[20]{}\mid {}\<[20E]%
\>[23]{}\Varid{dt}\leq \mathrm{0}{}\<[39]%
\>[39]{}\mathrel{=}\Varid{world}{}\<[E]%
\\[-0.3ex]%
\>[20]{}\mid {}\<[20E]%
\>[23]{}\Varid{dt}\mathbin{>}\Varid{gameRate}{}\<[39]%
\>[39]{}\mathrel{=}\Varid{gameStep}\;(\Varid{dt}\mathbin{-}\Varid{gameRate})\;(\Varid{step}\;\Varid{gameRate}\;\Varid{world}){}\<[E]%
\\[-0.3ex]%
\>[20]{}\mid {}\<[20E]%
\>[23]{}\Varid{otherwise}{}\<[39]%
\>[39]{}\mathrel{=}\Varid{step}\;\Varid{dt}\;\Varid{world}{}\<[E]%
\ColumnHook
\end{hscode}\resethooks
\caption{The complete client prediction code discussed in \cref{sec:committing}}
\label{fig:committing}
\end{figure}
\fi

\iflong
The main property of the code in \cref{fig:committing} is: No matter the interleaving of events from the various players, the result of \ensuremath{\Varid{currentState}} is the same. To increase our confidence that this property holds we used the QuickCheck library to randomly generate pairs of lists of events with monotonically increasing timestamps, considered all possible interleavings and checked that the resulting \ensuremath{\Conid{Log}\;\Varid{world}} data structure is identical.
\fi

\iflong
\section{Experiences and discussion}
\begin{wrapfigure}[11]{r}{0.4\linewidth}%
\vspace{-3em}
\fbox{\includegraphics[width=\linewidth]{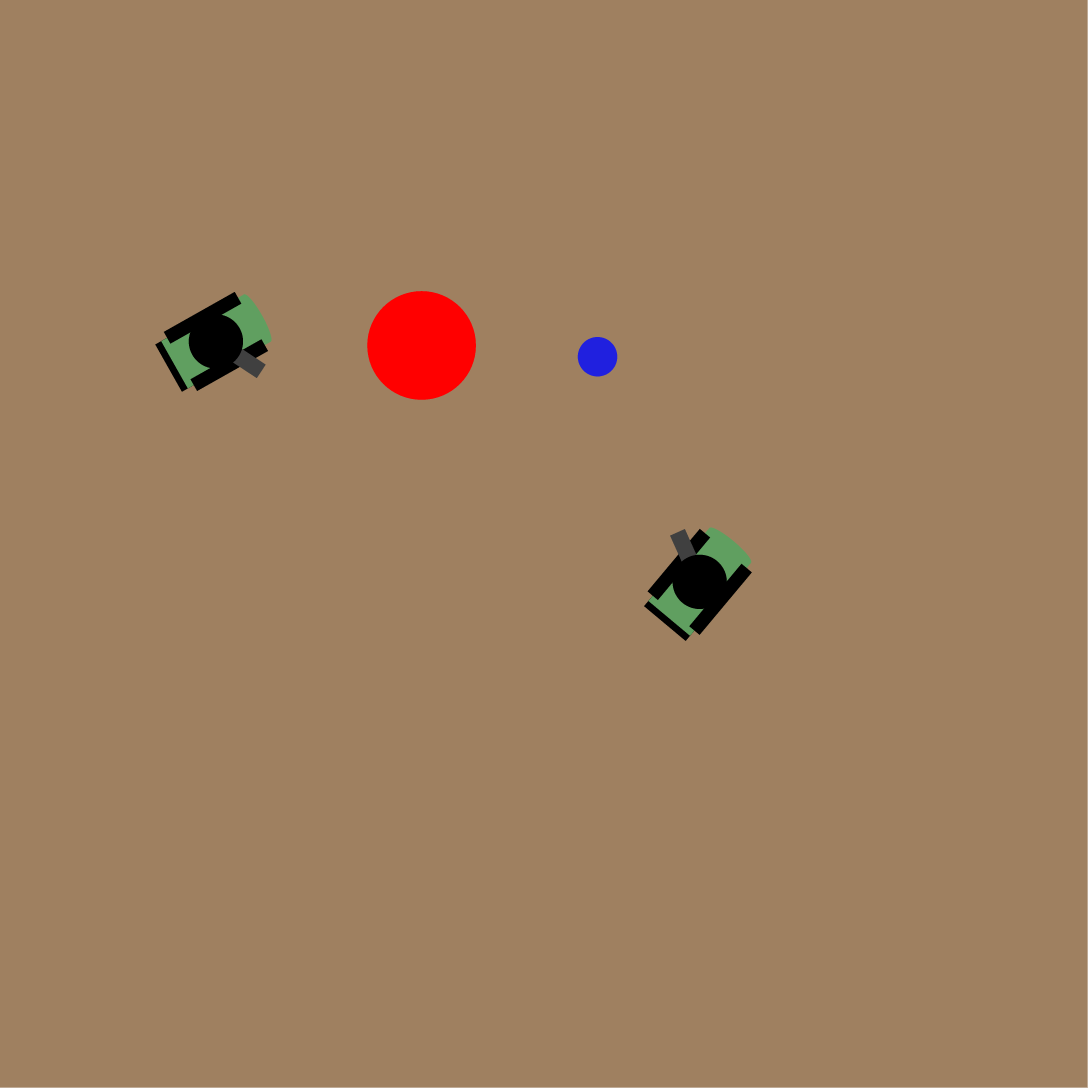}}
\caption{The tank game}
\label{fig:tank}
\end{wrapfigure}

The interface from \cref{sec:api} allows the creation of multi-user applications with great ease, and with the algorithms in \cref{sec:prediction}, CodeWorld can provide a smooth user experience. The reader may wonder, though, how well this works in practice, and what the drawbacks are for this approach.

\subsection{Early experience}
\label{sec:experience}

For a first practical evaluation of the system, the second author organized a stress test, involving four colleagues, a selection of games with different styles, and small prizes for winners.  During the event, participants play-tested the games, hoping to uncover any bugs or unexpected quirks of the format.  The games involved, which can be played at \url{https://code.world/gallery-icfp17.html}, include:
\begin{compactitem}
\item The Dot Grab game (\cref{fig:dotgrab}), which was originally written by a student as a single-computer interaction. Since the API for games is a straightforward extension of the one for interactions, it was trivial to make this game networked.
\item The game ``Snake" (\cref{fig:snake}), where a player has to move across the playing field while avoiding the other player's trails and the walls.
\item A tank game (\cref{fig:tank}) where each player steers a tank using the keyboard, aims using the mouse and fires bullets that explode after a certain time. Here the game evolves over time and manages a larger number of moving parts -- tanks, bullets, and explosions.
\end{compactitem}
Manual testing showed that the system is nicely responsive and that the artifacts due to network latency are noticeable, but not irritating. The system handled the more complex tank game well. A separate test using a high latency satellite connection remained playable, but with more pronounced latency-related artifacts, as expected.

We plan to introduce the API to students in the Spring semester of 2017.\jb{Let’s erase this or replace it with something more timeless.}
\else
\section{Discussion}

The interface from \cref{sec:api} allows the creation of multi-user applications with great ease, and with the algorithms in \cref{sec:prediction}, CodeWorld can provide a smooth user experience. But surely there are drawbacks and open problems worth discussing.%

\fi

\subsection{Floating point calculation}
\label{sec:floatingpoint}

A dominant concern in the implementation in \cref{sec:prediction} was to guarantee eventual consistency of all clients, so that game states would always converge over time. We achieve that requirement, on the assumption that the code passed to \ensuremath{\Varid{collaborationOf}} consists of pure functions. This result relies on a strong notion of pure function, though, which requires that outputs are predictable even between instances of the code running on different machines, operating systems, and runtime environments.  In this sense, even functions in Haskell may not always be pure!

A notable source of nondeterminism in Haskell is underspecified floating point operations. The \ensuremath{\Conid{Double}} type in Haskell is implementation-defined, and ``should cover IEEE double-precision" \citep{report2010}. Our interest is limited to the Haskell-to-JavaScript compiler GHCJS \citep{GHCJS}, which inherits the floating point operation semantics from JavaScript. The ECMA standard \citep{ecma} specifies a JavaScript number to be a ``double-precision 64-bit binary format IEEE 754-2008 value" -- which is luckily already a quite specific specification.  We are optimistic that the basic arithmetic operations are deterministic,
\iflong
and this optimism is supported by anecdotal reports from a game developer with Gas Powered Games \citep{emerson}
\begin{quote}
We have never had a problem with the IEEE standard across any PC CPU, AMD and Intel, with this approach. None of our [\ldots] customers have had problems with their machines either, and we are talking over 1 million customers here. We would have heard if there was a problem with the FPU not having the same results as replays or multi-player mode wouldn't work at all.
\end{quote}
\else
encouraged by reports from professional game developers \citep{emerson}.
\fi
However, transcendental functions (\ensuremath{\Varid{exp}}, \ensuremath{\Varid{sin}}, \ensuremath{\Varid{cos}}, \ensuremath{\Varid{log}}, etc.\@) are not completely specified by IEEE-754, and different browser/system combinations are allowed to yield slightly different results here.

We tested this with a double pendulum simulation, which makes heavy use of \ensuremath{\Varid{sin}} and \ensuremath{\Varid{cos}} in every simulation step. The double pendulum is a well-known example of a chaotic system, and we expect it to quickly magnify any divergence in state. Indeed, after running the program on two different browsers (Firefox and Chrome, on the same Linux machine) for several minutes, the simulations take different paths\iflong, confirming the worries about these functions\fi.

If, however, we use a custom implementation of \ensuremath{\Varid{sin}} -- based on a quadratic curve approximation -- the simulation runs consistently.
\iflong
We tested this variant on multiple JavaScript engines (Chrome, Firefox, and Microsoft Edge), on different operating systems (Windows, Linux, Android, and ChromeOS) and on different CPUs (Intel and ARM),
\else
We tested this variant on multiple JavaScript engines, OSs and CPUs,
\fi
and did not uncover any more consistency issues.  The tests confirm again that, apart from inconsistent implementations of transcendental functions, basic floating point operations are reliably deterministic in practice.

We can deploy a fix to transcendental functions in two ways. In CodeWorld's educational mode, where we have implemented a custom standard library, it is easy to just substitute new implementations of these functions. In the plain Haskell variant, however, we would like to allow the programmer to make use of existing libraries, which may use standard floating point functions. To achieve this, we can instead replace these operations at the JavaScript level, ensuring that even third-party Haskell libraries are deterministic.

\iflong
In the future, we also plan to automate checks for synchronization problems like this. We cannot directly compare program states in our implementation, since they are of arbitrary type.  However, we can compare the generated pictures -- or a hash thereof -- to achieve essentially the same effect.
\fi

\subsection{Interpolating the effects of delayed messages}
\label{sec:smoothing}

\begin{figure}
\centering
\begin{tikzpicture}[xscale=2,yscale=0.9]
\makeatletter
\newcommand\currentcoordinate{\the\tikz@lastxsaved,\the\tikz@lastysaved}
\makeatother

\def\s{0.5}
\def\E{5}

\def\A{0.8}
\def\Ar{1.2}
\def\As{1.5}
\def\B{1.6}
\def\Br{2.1}
\def\Bs{2.4}
\def\C{3.4}
\def\Cr{4.15}
\def\Cs{4.45}
\def\D{3.85}
\def\Dr{4.4}
\def\Ds{4.7}

\fill[color=gray!50] (0,0) -- (\E,0) -- (\E,\E*\s) -- cycle;
\begin{scope}
\clip                (0,0) -- (\E,0) -- (\E,\E*\s) -- cycle;
\foreach \x in {0,...,5} {
\draw (0,\x/3*\s) -- (\E,\x/3*\s);
}
\end{scope}

\fill[color=green]   (\A,\A*\s) -- (\E,\A*\s) -- (\E,\E*\s) -- cycle;
\begin{scope}
\clip                (\A,\A*\s) -- (\E,\A*\s) -- (\E,\E*\s) -- cycle;
\foreach \x in {0,...,5} {
\draw ($(0,\A*\s)+(0,\x/3*\s)$) -- ($(\E,\A*\s)+(0,\x/3*\s)$);
}
\end{scope}

\fill[color=red]     (\Bs,\B*\s) -- (\E,\B*\s) -- (\E,\E*\s) -- (\Br,\Br*\s) -- cycle;
\begin{scope}
\clip                (\Bs,\B*\s) -- (\E,\B*\s) -- (\E,\E*\s) -- (\Br,\Br*\s) -- cycle;
\foreach \x in {0,...,5} {
\draw ($(\Br,\Br*\s)+(0,\x/3*\s)$) -- ($(\Bs,\B*\s)+(0,\x/3*\s)$) -- ($(\E,\B*\s)+(0,\x/3*\s)$);
}
\end{scope}

\fill[color=yellow]  (\C,\C*\s) -- (\E,\C*\s) -- (\E,\E*\s) -- cycle;
\begin{scope}
\clip                (\C,\C*\s) -- (\E,\C*\s) -- (\E,\E*\s) -- cycle;
\foreach \x in {0,...,5} {
\draw ($(0,\C*\s)+(0,\x/3*\s)$) -- ($(\E,\C*\s)+(0,\x/3*\s)$);
}
\end{scope}

\fill[color=blue!60] (\Ds,\D*\s) -- (\E,\D*\s) -- (\E,\E*\s) -- (\Dr,\Dr*\s) -- cycle;
\begin{scope}
\clip                (\Ds,\D*\s) -- (\E,\D*\s) -- (\E,\E*\s) -- (\Dr,\Dr*\s) -- cycle;
\foreach \x in {0,...,5} {
\draw ($(\Dr,\Dr*\s)+(0,\x/3*\s)$) -- ($(\Ds,\D*\s)+(0,\x/3*\s)$) -- ($(\E,\D*\s)+(0,\x/3*\s)$);
}
\end{scope}

\draw[->] (0,0) node [left] {Player 1:} -- (\E,0) -- ++(0.25,0) node [below] {$t$};
\draw (0,0) -- (\E,\E*\s);

\node[inner sep=1pt, fill, circle] (A)  at (\A, 0) {};
\node[inner sep=1pt, fill, circle] (Br) at (\Br,0) {};
\node[inner sep=1pt, fill, circle] (Bs) at (\Bs,0) {};
\node[inner sep=1pt, fill, circle] (C)  at (\C ,0) {};
\node[inner sep=1pt, fill, circle] (Dr) at (\Dr,0) {};
\node[inner sep=1pt, fill, circle] (Ds) at (\Ds,0) {};

\begin{scope}[shift={(0,-6*\s)}]

\fill[color=gray!50] (0,0) -- (\E,0) -- (\E,\E*\s) -- cycle;
\begin{scope}
\clip                (0,0) -- (\E,0) -- (\E,\E*\s) -- cycle;
\foreach \x in {0,...,5} {
\draw (0,\x/3*\s) -- (\E,\x/3*\s);
}
\end{scope}

\fill[color=green]   (\As,\A*\s) -- (\E,\A*\s) -- (\E,\E*\s) -- (\Ar,\Ar*\s) -- cycle;
\begin{scope}
\clip                (\As,\A*\s) -- (\E,\A*\s) -- (\E,\E*\s) -- (\Ar,\Ar*\s) -- cycle;
\foreach \x in {0,...,5} {
\draw ($(\Ar,\Ar*\s)+(0,\x/3*\s)$) -- ($(\As,\A*\s)+(0,\x/3*\s)$) -- ($(\E,\A*\s)+(0,\x/3*\s)$);
}
\end{scope}

\fill[color=red]     (\B,\B*\s) -- (\E,\B*\s) -- (\E,\E*\s) -- cycle;
\begin{scope}
\clip                (\B,\B*\s) -- (\E,\B*\s) -- (\E,\E*\s) -- cycle;
\foreach \x in {0,...,7} {
\draw ($(0,\B*\s)+(0,\x/3*\s)$) -- ($(\E,\B*\s)+(0,\x/3*\s)$);
}
\end{scope}

\fill[color=yellow]   (\Cs,\C*\s) -- (\E,\C*\s) -- (\E,\E*\s) -- (\Cr,\Cr*\s) -- cycle;
\begin{scope}
\clip                (\Cs,\C*\s) -- (\E,\C*\s) -- (\E,\E*\s) -- (\Cr,\Cr*\s) -- cycle;
\foreach \x in {0,...,5} {
\draw ($(\Cr,\Cr*\s)+(0,\x/3*\s)$) -- ($(\Cs,\C*\s)+(0,\x/3*\s)$) -- ($(\E,\C*\s)+(0,\x/3*\s)$);
}
\end{scope}

\fill[color=blue!60] (\D,\D*\s) -- (\E,\D*\s) -- (\E,\E*\s) -- cycle;
\begin{scope}
\clip                (\D,\D*\s) -- (\E,\D*\s) -- (\E,\E*\s) -- cycle;
\foreach \x in {0,...,5} {
\draw ($(0,\D*\s)+(0,\x/3*\s)$) -- ($(\E,\D*\s)+(0,\x/3*\s)$);
}
\end{scope}

\def\CD{(\Cr+(\Cs-\Cr)/(\Cr-\C)*(\Cr-\D))}
\fill[color=yellow]   (\Cr,\Cr*\s) -- ({\CD},\D*\s) -- ({\CD},{\CD*\s})-- cycle;
\begin{scope}
\clip                 (\Cr,\Cr*\s) -- ({\CD},\D*\s) -- ({\CD},{\CD*\s})-- cycle;
\foreach \x in {0,...,5} {
\draw ($(\Cr,\Cr*\s)+(0,\x/3*\s)$) -- ($({\CD},\D*\s)+(0,\x/3*\s)$);
}
\end{scope}

\draw[->] (0,0) node [left] {Player 2:} -- (\E,0) -- ++(0.25,0) node [below] {$t$};

\draw (0,0) -- (\E,\E*\s);

\node[inner sep=1pt, fill, circle] (Ar) at (\Ar,0) {};
\node[inner sep=1pt, fill, circle] (As) at (\As,0) {};
\node[inner sep=1pt, fill, circle] (B)  at (\B, 0) {};
\node[inner sep=1pt, fill, circle] (Cr) at (\Cr,0) {};
\node[inner sep=1pt, fill, circle] (Cs) at (\Cs,0) {};
\node[inner sep=1pt, fill, circle] (D)  at (\D, 0) {};
\end{scope}

\draw[dashed, ->] (A) -- (Ar);
\draw[dashed, ->] (B) -- (Br);
\draw[dashed, ->] (C) -- (Cr);
\draw[dashed, ->] (D) -- (Dr);

\draw[->] (Ar) edge[bend right] (As);
\draw[->] (Br) edge[bend right] (Bs);
\draw[->] (Cr) edge[bend right] (Cs);
\draw[->] (Dr) edge[bend right] (Ds);

\node[below,fill=white,inner sep=0pt,outer sep=4pt] at (A) {\keystroke{1}};
\node[below,fill=white,inner sep=0pt,outer sep=4pt] at (B) {\keystroke{2}};
\node[below,fill=white,inner sep=0pt,outer sep=4pt] at (C) {\keystroke{3}};
\node[below,fill=white,inner sep=0pt,outer sep=4pt] at (D) {\keystroke{4}};

\end{tikzpicture}
\caption{Smoothing the effect of late events}
\label{fig:smoothevolution}
\end{figure}
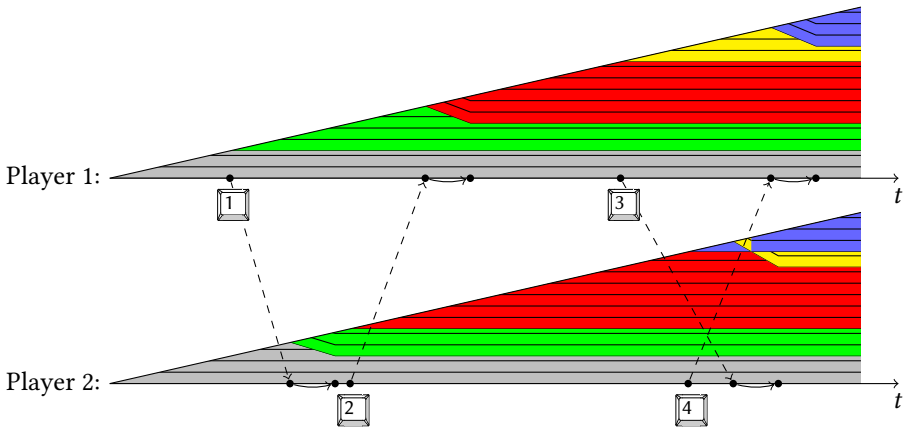
Another trick in the game programming toolbox is interpolation to smooth out artifacts that result from corrections to the game state. These artifacts can be clearly seen in \cref{fig:evolution}: The moment the message \keystroke{2} reaches the first player, the top segment of the growing column abruptly changes from green to red.
\iflong
Similarly, in a game like the tank-fighting game in \cref{fig:tank}, an opponent can appear to teleport to a new location.
\else
Similarly, in a first-person shooter, an opponent can appear to teleport to a new location.
\fi
In this situation, many games would instead interpolate the position smoothly over a fraction of a second. This can introduce new anomalies of its own, \iflong such as characters passing through walls, or tanks moving sideways, \fi but in most cases, it is hoped the result will appear more realistic than the alternative.

\iflong
By providing an API that is completely abstract in the game state, it seems that we have shut the door on implementing this trick.
We lack the ability to look inside the state and adjust positions. Surprisingly, though, a form of interpolation is possible.
All that is needed is a sort of change of coordinates.
While we cannot interpolate in space, we can interpolate in time!
\else
It seems that our API is too abstract to allow us to implement this trick, as we cannot look inside the state to adjust positions. But although we cannot interpolate in space, we can interpolate in time!
\fi
When a delayed event arrives, we initially treat it as if its timestamp is ``now" and then slide it backward in time over a short interpolation period until it reaches its actual time.

Usually, the \ensuremath{\Varid{step}} function is approximately continuous, and as a result, moving an event backwards in time gives a smooth interpolation in the state as well. This can be seen in \cref{fig:smoothevolution}: After the message \keystroke{2} arrives at Player 1, the column smoothly changes its color from green to red, from the tip downwards, until the correct state is reached.  Like all interpolation, though, anomalies can still happen.  This scheme introduces abrupt artifacts as we slide a delayed message past another event with a non-commuting effect. In \cref{fig:smoothevolution} the second player smoothly integrates the delayed \keystroke{3} message, and the top of the column changes color from blue to yellow. But the moment this event is pushed before the local event \keystroke{4}, the column abruptly changes its color back to blue.

This is an elegant trick to recover the ability to do interpolation.  However, it is not clear if interpolation is always the best experience, and a jerky, abrupt update may be preferred for certain games.

\subsection{Irreversible updates}

In some cases, the visual artifacts due to delayed messages, whether smooth or jerky, pose a serious problem. Consider, for example, a card game in which both players click to draw cards from the same deck.  Suppose player~1 clicks to draw a card first, but the message from player~1 to player~2 arrives after player~2 clicks as well.  For a brief moment before the message is received, player~2 sees the top card, even though it ultimately ends up in the first player's hand! This is an example of a case where eventual consistency in the game state is not good enough.

This problem is hard to avoid, given our constraints and the third requirement of responding immediately to local events.  It can be mitigated by the game programmer, by adding a short delay before major events such as those that reveal secrets. The delay can sometimes be creatively hidden by animations or effects. This trick dodges the problem as long as network latency is shorter than this delay, but it provides no guarantee. A complete solution to this problem must involve the programmer in a way that is undesirable in our setting, since only the programmer understands which state changes represent a significant enough event to postpone.

\iflong
\subsection{Lock-step simulation and CRDTs}

Our approach to lock-step simulation may remind some readers of conflict-free replicated data types (CRDTs), introduced by \citet{shapiro:inria-00555588} as a lightweight approach to providing strong consistency guarantees in distributed systems, even in the face of network failure, partition or out-of-order event delivery. These data types come in two forms: ``convergent" replicated data types (CvRDTs) are based on transmitting state directly, while ``commutative" replicated data types (CmRDTs), are based on transmitting operations that act on that state. Despite the similarity, our game state does not form a CmRDT, as these require that update operations on the game state are commutative.  This limits the types of data that can used in such an approach and is inconsistent with our first requirement of supporting arbitrary game state.

We find, however, that type \ensuremath{\Conid{Log}} type defined in \cref{fig:committing} forms a CmRDT\@. The \ensuremath{\Varid{addEvent}} events from different players commute, as both just add the event to the set. The theory of CRDTs hence provides another argument that the resulting game state is eventually consistent (in fact, strongly so).
\fi

\section{Conclusions}

By implementing lock-step simulation with client prediction generically in the educational programming environment CodeWorld, we have demonstrated once more that that pure functional programming excels at abstraction and modularity. In addition, this work will directly support the education of our next generation of programmers.

\begin{acks} 
  The first author has been supported by the
  \grantsponsor{GS100000001}{National Science
    Foundation}{http://dx.doi.org/10.13039/100000001} under Grant
  No.~\grantnum{GS100000001}{CCF-1319880} and Grant
  No.~\grantnum{GS100000001}{14-519}.
  We thank Stephanie Weirich, Zach Kost-Smith, Sterling Stein and Justin Hsu for helpful comments on a draft of this paper, as well as the reviewers for their comments.
\end{acks}

\bibliography{codeworld-icfp17}

\end{document}